# Timing of Millisecond Pulsars in NGC 6752. III. On the Presence of Nonluminous Matter in the Cluster's Core

A. Corongiu[1], A. Ridolfi[1,2], F. Abbate[1,2], M. Bailes[3,4], A. Possenti[1], M. Geyer[5,6], R. N. Manchester[7], M. Kramer[2,8], P. C. C. Freire[2], M. Burgay[1], S. Buchner[5], and F. Camilo[5]

[1] INAF—Osservatorio Astronomico di Cagliari, Via Della Scienza 5, I-09047 Selargius, Italy; alessandro.corongiu@inaf.it
[2] Max-Planck-Institut für Radioastronomie, Auf dem Hügel 69, D-53121 Bonn, Germany
[3] Centre for Astrophysics and Supercomputing, Swinburne University of Technology, PO Box 218, Hawthorn, Vic, 3122, Australia
[4] The ARC Centre of Excellence for Gravitational Wave Discovery (OzGrav), Mail H29, Swinburne University Of Technology, PO Box 218, Hawthorn, Vic 3122, Australia
[5] South African Radio Astronomy Observatory, 2 Fir Street, Black River Park, Observatory 7925, South Africa
[6] Department of Astronomy, University of Cape Town, Rondebosch, Cape Town, 7700, South Africa
[7] Australia Telescope National Facility, CSIRO, Space and Astronomy, Epping NSW 1710, Australia
[8] Jodrell Bank Centre for Astrophysics, School of Physics and Astronomy, Univ. of Manchester, Manchester, M13 9PL, UK
*Received 2023 August 1; revised 2024 June 25; accepted 2024 July 1; published 2024 September 10*

## Abstract

Millisecond pulsars are subject to accelerations in globular clusters (GCs) that manifest themselves in both the first and second spin period time derivatives, and can be used to explore the mass distribution of the potentials they inhabit. Here we report on over 20 yr of pulsar timing observations of five millisecond radio pulsars in the core of the core-collapse GC NGC 6752 with the Parkes (Murriyang) and MeerKAT radio telescopes, which have allowed us to measure the proper motions, positions, and first and second time derivatives of the pulsars. The pulsar timing parameters indicate that all the pulsars in the core experience accelerations and jerks that can be explained only if an amount of nonluminous mass of at least $2.56 \times 10^3 M_\odot$ is present in the core of NGC 6752. On the other hand, our studies highly disfavor the presence of an intermediate-mass black hole at the center of the cluster, with a mass equal to or greater than $\sim 3000 M_\odot$.

*Unified Astronomy Thesaurus concepts:* Millisecond pulsars (1062); Pulsars (1306); Globular star clusters (656)

## 1. Introduction

Globular clusters (GCs) are among the oldest and densest structures in a galaxy and often contain several $\times 10^5$ stars in spherical structures with radii of just a few tens of parsecs. In recent decades, more and more powerful optical telescopes have been investigating these systems, providing new insights into their mass distribution and internal structure (see, e.g., Baumgardt & Hilker 2018; Leitinger et al. 2023, and references therein). Today accurate censuses of their stellar populations can be obtained via optical data, and their structure, dynamical status, and evolution can be explored with $N$-body simulations. While their macroscopic properties are increasingly well understood, their dense cores remain difficult to probe. Even high spatial resolution instruments like GAIA (Gaia Collaboration et al. 2016) struggle to resolve stars in the central regions due to the high number densities and distances. Another reason is the presence in the central regions of high-mass nonluminous stellar remnants like neutron stars (NS) and possibly stellar mass black holes (Bahcall & Wolf 1976, Bahcall & Wolf 1977), and also massive underluminous white dwarfs (e.g., Vitral et al. 2023) which have migrated to the clusters' cores, thus leaving the lightest objects at larger distances from the center due to mass segregation. This scenario finds confirmation via the observed spatial distribution of millisecond pulsars (MSP) in GCs, which are on average among the heaviest objects in a cluster and usually are located near the cores. The pulsars located outside the core have been postulated to be the result of multibody ejection mechanisms, some involving the presence of an intermediate-mass black hole (IMBH) of several tens of solar masses (e.g., Colpi et al. 2002), located near the cluster's center of gravity. In recent works, instead, it has been argued that three-body interactions involving stellar mass objects are a viable mechanism for expelling binary MSPs from the cluster core (see, e.g., Leigh et al. 2024).

The MSPs in the GC NGC 6752 show characteristics that are typical of the populations of core-collapsed GCs: a large predominance of isolated pulsars (for other examples, see the pulsar population of M15, Anderson 1993, NGC 6624, Abbate et al. 2022 and NGC 6517, Pan et al. 2021) and a much larger fraction of pulsars at large distances from the core. As discussed in detail by Verbunt & Freire (2014), these facts are likely causally related: the large number of stellar interactions per binary in the core region—to which NSs migrate via dynamical friction—implies that many MSP binaries are disrupted in three-body encounters, with many being ejected from the cluster, and with several others barely staying bound to it. The ejections of NSs are more likely if they are involved in interactions with more massive stellar remnants.

However, even in comparison with the pulsar populations of other core-collapsed GCs, the population of NGC 6752 is extreme in these respects. Its peculiarities are hardly explainable without invoking the presence of a nonnegligible amount of under- or nonluminous (hereafter, we name all of that as nonluminous) mass in the core, under the form of massive remnants. PSR J1910–5959A (hereafter PSRA, D'Amico et al. 2001), the first pulsar discovered in this cluster and its only binary pulsar known to date, has the largest offset from the center of its host GC in core radius units, namely $\sim 58\, r_c$ (this fact has cast doubts on the association, which has, however, been demonstrated recently by Corongiu et al. 2023) and angular units (∼6.′3); while PSR J1910−5959C (hereafter PSRC, D'Amico et al. 2002, hereafter Paper I) has the third angular offset in core radius units, namely ∼23 $r_c$, and fourth in angular units (∼2.′6[9]). For these two pulsars, recent episodes of ejection from the core are a very likely possibility, since dynamical friction should have taken them back to the core in less than ∼1 Gyr (Colpi et al. 2002). The other known pulsars, namely PSR J1910–5959B, PSR J1910–5959D, and PSR J1910–5959E (hereafter PSRB, PSRD, and PSRE respectively, Paper I), and the recently discovered PSR J1910−5959F (hereafter PSRF, Ridolfi et al. 2021) are located in the core of NGC 6752. We will refer to these four pulsars as the *core pulsars*.

In Paper I it was shown that the measured spin period derivatives of PSRB, PSRD, and PSRE are dominated by the acceleration imparted to them by the cluster potential well, and at least $1.3 \times 10^4 M_\odot$ should be present in the form of nonluminous objects in the core. The measured spin period derivative of PSRF (Ridolfi et al. 2021) is of the same order of magnitude as the other three core pulsars, and also its value is likely due to the gravitational effect of nonluminous matter in the core of NGC 6752.

In this work, we report updated timing solutions for PSRB, PSRC, PSRD, PSRE, and PSRF[10] based on observations taken over ∼20 yr with the Parkes Murriyang and MeerKAT radio telescopes, and investigate the total mass and distribution of the nonluminous matter in the cluster core. This paper is organized as follows: In Section 2 we describe the observations and the data analysis, and in Section 3 we comment on the proper motions of the core pulsars. In Section 4 we place constraints on the nonluminous matter in the cluster core, and in Section 5 we discuss our results in terms of the central mass-to-light ratio. In Section 6 we comment on the main sources of uncertainty in our analysis, and we review our numerical results. Finally, in Section 7 we summarize our findings.

## 2. Observations and Data Analysis

### 2.1. Parkes Observations

The GC NGC 6752 was observed with the Parkes Murriyang 64 m radio telescope, located in Australia, from 1999 September to 2016 March. Observations were carried out in the *L*-band with the Multibeam and the H−OH receivers, depending on availability, with a bandwidth of 256 MHz and using two orthogonal polarizations that were detected and summed before digitization. The Stokes *I* signal was processed with the 1 bit Analog Filterbank (AFB; see, e.g., D'Amico et al. 2001) digital signal processor (hereafter "backend") up to its decommissioning in 2012 September. The AFB acquired the detected signal at the central frequency of 1390 MHz, split the observed bandwidth into 512 frequency channels, and recorded the data with a sampling time of 80 μs (125 μs in some earlier observations). In all observations taken after 2012 September, and in some before, the data were recorded in search mode with the Pulsar Digital Filterbank 3 (PDFB3) and 4 (PDFB4; see, e.g., Manchester et al. 2013 and references therein). These backends acquired the detected signal at the central frequency of 1369 MHz, split the observed bandwidth in 512 (DFB3) or 1024 (DFB4) frequency channels, and recorded the data with a sampling time of 80 μs (125 μs in a few observations).

We reduced and analyzed the raw data with the same method used and explained in detail in Paper I and Corongiu et al. (2006, hereafter Paper II). Briefly, for each pulsar, we folded the data at the approximate topocentric period with the software package dspsr[11] (van Straten & Bailes 2011), as predicted by the ephemeris published in Paper II. We folded the data with a typical subintegration length of 1 minute and kept the same number of frequency channels as the raw data to aid in accurate dedispersion. The resulting reduced data, whose files are usually described as "timing archives," have the form of a matrix whose elements are the binned profile amplitudes observed in each frequency channel, one for each subintegration time, and some metadata (header) information.

### 2.2. MeerKAT Observations

The MeerKAT radio telescope (Jonas & MeerKAT Team 2016; Camilo et al. 2018), located in South Africa, is an array composed of 64 offset-Gregorian dishes, each equivalent to a circular aperture of approximately 13.5 m across. We used MeerKAT and its UHF-band (544–1088 MHz) and *L*-band (856–1712 MHz) receivers to observe NGC 6752 on several occasions between 2019 July and 2022 January, as part of the MeerTime[12] (Bailes et al. 2020) and the TRansients And PUlsars with MeerKAT (TRAPUM[13]; Stappers & Kramer 2016) Large Survey Projects. Each of these projects has its own backend, tailored to its main scientific goals. MeerTime, which is mostly devoted to precision pulsar timing and polarimetry, makes use of the Pulsar Timing User Supplied Equipment (PTUSE) backend. This is capable of acquiring up to four Nyquist-sampled tied-array beams with up to 4096 coherently dedispersed frequency channels with four Stokes parameters. The data were recorded as 8 bit search mode PSRFITS files with a time resolution of 9.57 μs at *L*-Band and 7.53 μs at UHF. Given that only pulsars A and C are located far from the core of the cluster, while all the others are less than 0.′′2 away from the nominal center of the cluster, three PTUSE beams (centered on pulsar A, on pulsar C, and on the nominal center of NGC 6752, respectively) were sufficient to record the signals from all the known pulsars. Also, being the DMs of the pulsars all in the range of 33.20–33.70 pc cm$^{-3}$, a single search mode file coherently dedispersed at a DM = 33.3 pc cm$^{-3}$ was sufficient to allow the redetection of all the pulsars without any loss of sensitivity. TRAPUM, on the other hand, is focused on the search for pulsars and radio transients, and uses two computing clusters. First, the Filterbanking BeamFormer User Supplied Equipment (FBFUSE) cluster applies the beamforming technique to combine the raw signals from all the antennas and synthesize hundreds (up to 288 for our NGC 6752 observations) tied-array beams on the sky (Barr 2018; Chen et al. 2021). These are then recorded as search mode "filterbank" files by the Accelerated Pulsar Search User Supplied Equipment (APSUSE) cluster: the observing band is recorded with a typical time resolution of ∼60–80 μs and is split into 4096 frequency channels. The fine channelization is particularly important in this case to remove the effects of

---

[9] See https://www3.mpifr-bonn.mpg.de/staff/pfreire/GCpsr.html for a complete list of all published pulsars in GCs.
[10] The timing solution for PSRA, based on the same observations, has recently been published by Corongiu et al. (2023).
[11] https://dspsr.sourceforge.net/
[12] http://www.meertime.org
[13] http://www.trapum.org

**Table 1**
List of the MeerKAT Observations of NGC 6752 Used for This Work

| Obs. ID | Project | Start MJD | Length (s) | Backend | $f_c$ (MHz) | $\Delta f$ (MHz) | CD DM (pc cm$^{-3}$) | $N_{\text{chan}}$ | $N_{\text{pol}}$ | $t_{\text{samp}}$ ($\mu$s) | $N_{\text{ant}}$ |
|---------|---------|-----------|------------|---------|-------------|------------------|----------------------|-------------------|------------------|----------------------------|------------------|
| 01L | MeerTime | 58666.786 | 9000 | PTUSE | 1284 | 642 | 33.7 | 768 | 4 | 9.57 | 61 |
| 02L | MeerTime | 58850.637 | 9000 | PTUSE | 1284 | 642 | 33.28 | 768 | 4 | 9.57 | 61 |
| 03L | MeerTime | 59059.953 | 7200 | PTUSE | 1284 | 856 | 33.29 | 4096 | 4 | 19.14 | 40 |
| 04L | TRAPUM | 59174.682 | 7200 | APSUSE | 1284 | 856 | ⋯ | 4096 | 1 | 76.56 | 56 |
|     |        | 59174.682 | 7200 | PTUSE | 1284 | 856 | 33.29 | 4096 | 4 | 19.14 | 42 |
| 05U | TRAPUM | 59236.368 | 7000 | APSUSE | 816 | 544 | ⋯ | 4096 | 1 | 60.24 | 56 |
|     |        | 59236.365 | 7200 | PTUSE | 816 | 544 | 33.29 | 4096 | 1 | 22.59 | 41 |
| 06U | TRAPUM | 59244.401 | 6950 | APSUSE | 816 | 544 | ⋯ | 4096 | 1 | 60.24 | 60 |
|     |        | 59244.398 | 7200 | PTUSE | 816 | 544 | 33.29 | 4096 | 1 | 7.53 | 41 |
| 07L | MeerTime | 59389.691 | 7200 | PTUSE | 1284 | 856 | 33.7 | 4096 | 1 | 9.57 | 61 |
|     |        | 59389.727 | 18400 | APSUSE | 1284 | 856 | ⋯ | 4096 | 1 | 76.56 | 60 |
| 08L | MeerTime | 59451.634 | 21600 | PTUSE | 1284 | 856 | 33.7 | 4096 | 1 | 9.57 | 58 |
|     |        | 59451.635 | 21600 | APSUSE | 1284 | 856 | ⋯ | 4096 | 1 | 76.56 | 56 |

**Note.** The symbols stand for the following: $f_c$, central frequency; $\Delta f$, observing bandwidth; CD DM: dispersion measure used for coherent dedispersion; $N_{\text{chan}}$, number of frequency channels; $N_{\text{pol}}$, number of Stokes parameters; $t_{\text{samp}}$, sampling time; $N_{\text{ant}}$, number of antennas.

interstellar dispersion, as APSUSE does not benefit from coherent dedispersion, due to computational constraints. The majority of the MeerKAT observations of NGC 6752 were carried out using PTUSE and APSUSE simultaneously, so as to take advantage of their complementary characteristics. The exact used setups depended on the main scientific purpose of each observing session, as well as on the availability of new observing modes (such as the possibility of recording more than a single beam with PTUSE, or to record a different number of channels between PTUSE and APSUSE during a simultaneous MeerTime+TRAPUM session) which were gradually implemented over the course of these projects. We excluded the APSUSE data for PSRB, because the time stamp was not correctly recorded in the corresponding data files. We report all the MeerKAT observations of NGC 6752 used in this paper, along with their configurations, in Table 1.

### 2.3. ToA Extraction and Timing

We extracted the pulse times of arrival (ToAs) with the routines of the software suite psrchive (Hotan et al. 2004), by coherently adding in phase the profiles in each archive with respect to subintegrations and channels, and convolving them with a high signal-to-noise ratio (S/N) template obtained by summing in phase the observed profiles with the best S/N. For each pulsar, we produced a specific template for each telescope-backend-observing mode combination. The pulsars were often subject to strong scintillation, due to the ionized interstellar medium, which produces random variations of the pulse S/N with respect to both time and frequency, correlated on some timescale. For this cluster the timescale is such that amplitude variations occur not only from one observation to another, but also within single observations, and they are so dramatic that pulses can be very bright at some moments and/ or in some frequency subbands, and completely undetectable in others. For these reasons, we could not determine a single optimal integration length and subband frequency width along which the profiles in the archive should always be summed; instead, we visually inspected every single archive to determine the best time interval(s) and frequency band(s) where pulses were evaluated as detections. Whenever possible, we extracted more than one ToA from a single observation, either by decimating in time, frequency, or both.

We determined the instrumental time jump between the different backends used for Parkes observations, and between the Parkes and MeerKAT data sets as a whole. We determined these jumps for each pulsar separately, since they are not due to technical reasons only, but mainly because templates differ, from one pulsar to another, in shape, displacement, and time extent. We used the ToAs obtained from the Parkes AFB backend as reference (i.e., we determined the time jumps with respect to), since for all pulsars these sets span the largest time interval and contain the largest number of ToAs. At first we considered the ToAs for each telescope separately, and fitted them against the best timing model for each set. We took the resulting $\chi^2$ and multiplied the ToA uncertainties by its square root. Once this operation is performed, a new fit on the same ToAs against the same timing model should return a reduced $\chi^2 = 1$. We then combined the two telescope's ToAs and fitted them for a single time jump alone, using Parkes ToAs as the reference. Finally, we proceeded with the fit against the pulsar timing model, still allowing this time jump to vary for correctly estimating the uncertainties in the timing model's parameters.

We fitted the obtained ToAs against the rotational and astrometric parameters with the pulsar timing software TEMPO2 (Hobbs et al. 2006), using the DE430 solar system ephemeris from the NASA Jet Propulsion Laboratory[14] for barycentering. Table 2 reports our timing results. The values for each parameter in the timing model are reported with the nominal TEMPO2 1$\sigma$ uncertainty on the last significant digit (in parentheses). The main improvement in the timing solutions with respect to Paper II is the determination of the second derivative of the spin period for all pulsars, the proper motions for PSRB, D, and E, and the refinement of the latter for PSRF. Parameters whose measurement has already been reported in previous works are now determined with a similar or higher precision. Figure 1 displays the corresponding timing residuals. The amplitude of the plotted error bars is equal to the measured one for each ToA, multiplied by the amount necessary to obtain $\chi^2 = 1$ in the preliminary fit of each telescope data set separately (see above). No clear trend is evident in the

---
[14] https://www.jpl.nasa.gov, see the web page https://ssd.jpl.nasa.gov/planets/orbits.html.

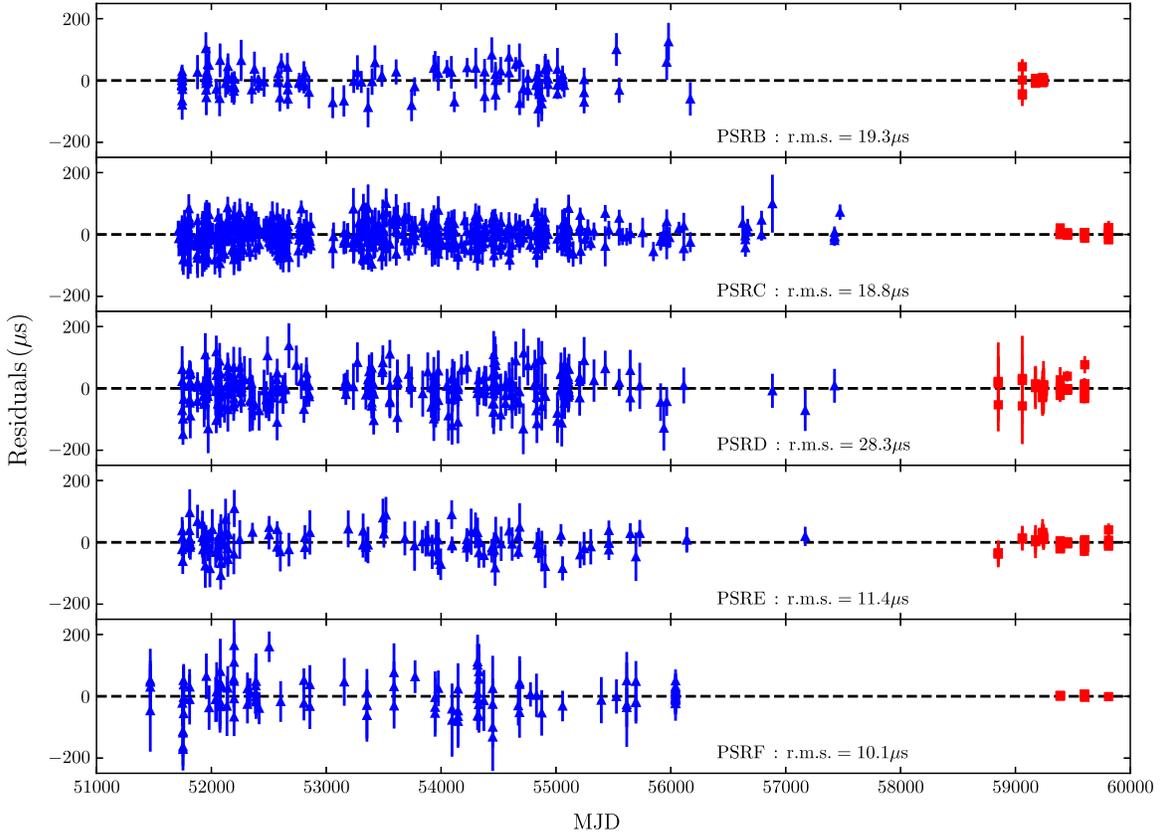

**Figure 1.** Timing residuals vs. MJD. ToAs obtained with Parkes and MeerKAT are plotted with blue triangles and red squares, respectively. If not visible, error bars are smaller than the marking symbol. Units and scales are the same for all pulsars for an immediate comparison of their timing rms and noise level. Pulsar names and rms values are indicated near the bottom right corner of the dedicated panel.

Table 2
Measured and Derived Parameters for the Pulsars in NGC 6752

| Parameter | PSR B | PSR C | PSR D | PSR E | PSR F |
|---|---|---|---|---|---|
| Measured parameters | | | | | |
| R.A.[a] (J2000) | 19:10:52.05567(26) | 19:11:05.55529(13) | 19:10:52.41706(26) | 19:10:52.15642(31) | 19:10:52.0628(12) |
| Decl.[a] (J2000) | −59:59:00.863(3) | −60:00:59.7017(13) | −59:59:05.4724(26) | −59:59:02.0846(36) | −59:59:09.2699(42) |
| $\mu_\alpha \cos\delta$ (mas yr$^{-1}$) | −2.37(24) | −3.15(8) | −3.36(18) | −3.09(12) | −3.76(66) |
| $\mu_\delta$ (mas yr$^{-1}$) | −4.03(46) | −3.76(9) | −3.69(22) | −4.26(19) | −4.54(41) |
| $\nu$ (Hz) | 119.648732845014(14) | 189.489871070457(10) | 110.677191984295(11) | 218.733857589954(24) | 117.848178012089(51) |
| $\dot\nu$ (10$^{-14}$ Hz s$^{-1}$) | 1.131889(22) | −0.007728(9) | −1.18112(1) | 2.07904(2) | −1.02912(3) |
| $\ddot\nu$ (10$^{-26}$ Hz s$^{-2}$) | −6.46(15) | −0.015(29) | −1.742(44) | −14.93(6) | −8.6(1) |
| DM (pc cm$^{-3}$) | 33.2947(12) | 33.2817(4) | 33.297(1) | 33.32505(54) | 33.22246(51) |
| Epoch[b] (MJD) | 52,000 | 51,910 | 51,910 | 51,910 | 58700 |
| MJD range | 51,741–59,244 | 51,710–59,810 | 51,745–59,604 | 51,744–59,810 | 51,468–59,810 |
| Number of TOAs | 149 | 661 | 350 | 172 | 130 |
| Residuals rms ($\mu$s) | 19.3 | 18.8 | 28.3 | 11.4 | 10.1 |
| Derived parameters | | | | | |
| $P$ (ms) | 8.3577985008445(10) | 5.2773269323096(3) | 9.0352852477672(9) | 4.5717659397506(4) | 8.485493936932(1) |
| $\dot P$ (10$^{-19}$ s s$^{-1}$) | −7.90656(15) | 0.02152(3) | 9.64225(9) | −4.34541(45) | 7.41005(25) |
| $\ddot P$ (10$^{-30}$ s s$^{-2}$) | 4.5(1) | 0.0043(82) | 1.422(36) | 3.120(13) | 6.200(73) |
| $\mu$ (mas yr$^{-1}$) | 4.67(41) | 4.91(9) | 4.99(20) | 5.26(17) | 5.89(52) |
| Pos. angle (degrees) | 210.5(38) | 219.9(10) | 222.3(23) | 216.0(16) | 219.6(56) |
| Offset[c] (arcmin) | 0.063 | 2.56 | 0.049 | 0.045 | 0.077 |

**Notes.**
[a] Units of R.A. are hours, minutes, and seconds, and units of decl. are degrees, arcminutes, and arcseconds.
[b] Reference epoch for both the spin period and the position.
[c] The offset of the pulsars is calculated with respect to the position of the cluster's center of gravity reported by Ferraro et al. (2003; see Table 3).

**Table 3**
Positional, Kinematical, and Structural Parameters for NGC 6752 Published in Literature and Used in This Work

| Parameter | Symbol | Value | References |
|---|---|---|---|
| R.A. (hh:mm:ss.ss, J2000) | $\alpha$ | 19:10:52.04 | (a) |
| decl. (dd:mm:ss.ss, J2000) | $\delta$ | −59:59:04.64 | (a) |
| Position Epoch (MJD) | ⋯ | 51990 | (a) |
| Proper motion in R.A. (mas yr$^{-1}$) | $\mu_\alpha \cos\delta$ | $-3.161 \pm 0.022$ | (b) |
| Proper motion in Decl. (mas yr$^{-1}$) | $\mu_\delta$ | $-4.027 \pm 0.022$ | (b) |
| Distance (kpc) | $D$ | $4.125 \pm 0.040$ | (c) |
| Total mass ($10^5 M_\odot$) | $M_{\rm TOT}$ | $2.76 \pm 0.04$ | (d) |
| Core radius (pc) | $r_c$ | 0.13 | (d) |
| Central surface brightness ($10^4 L_\odot$ pc$^{-2}$) | $\Sigma_V$ | 4.025 | (e) |
| Central density ($M_\odot$ pc$^{-3}$) | $\rho_0$ | $2.4 \times 10^5$ | (f) |
| Central escape velocity (km s$^{-1}$) | $V_{\rm esc,c}$ | 34.5 | (f) |
| Central proper motion dispersion (mas yr$^{-1}$) | $\sigma_\mu$ | $0.436 \pm 0.009$ | (g) |
| Central velocity dispersion (km s$^{-1}$) | $\sigma_0$ | $8.5 \pm 0.2$ | (g) derived |

**References.** (a) Ferraro et al. (2003); (b) Vasiliev & Baumgardt (2021); (c) Baumgardt & Vasiliev (2021); (d) Globular clusters online catalog https://people.smp.uq.edu.au/HolgerBaumgardt/globular/, v.3, see Hilker et al. (2020); (e) Noyola & Gebhardt (2006); (f) Baumgardt & Hilker (2018); (g) Libralato et al. (2022).

residuals of any pulsar, thus indicating that the adopted timing models adequately describe the available data sets.

## 3. Proper Motions of the Core Pulsars and the Central Velocity Dispersion

The pulsars in the core of NGC 6752 have excellent proper motion determinations, namely with a significance of $11\sigma$, $25\sigma$, $31\sigma$, and $11\sigma$ for PSRB, PSRD, PSRE, and PSRF in alphabetical order, respectively. The average motion of the pulsars results in $\langle\mu_\alpha \cos\delta\rangle = -3.15 \pm 0.51$ mas yr$^{-1}$ (quoted uncertainties are everywhere at the 1$\sigma$ level unless explicitly indicated) and $\langle\mu_\delta\rangle = -4.13 \pm 0.31$ mas yr$^{-1}$, in excellent agreement with the cluster motion $\mu_\alpha \cos\delta = -3.162 \pm 0.023$ mas yr$^{-1}$ and $\mu_\delta = -4.028 \pm 0.022$ mas yr$^{-1}$, obtained from GAIA optical observations (all cluster's parameters picked up from literature and used in this work are presented in Table 3 with all necessary references, unless explicitly indicated). At the cluster distance of $4.125 \pm 0.040$ kpc the relative projected transverse velocity of each pulsar with respect to the cluster center, for which the GAIA proper motion is assumed, are $15.5 \pm 4.7$, $7.7 \pm 4.1$, $4.8 \pm 3.6$, and $15.4 \pm 11.1$ km s$^{-1}$ respectively, thus implying that (barring a large radial velocity component) all pulsars are compatible with being bound to the cluster core, since the escape velocity from the core is $V_{\rm esc} = 34.5$ km s$^{-1}$.

We estimated the 1D velocity dispersion of the core pulsars, by calculating the standard deviation of the proper motion components in R.A. and decl. separately, then averaging the values in the two directions. We obtain $\sigma_{\rm V,PSRs} = 8.0 \pm 1.9$ km s$^{-1}$. This value is compatible with the value $\sigma_0 = 8.5 \pm 0.2$ km s$^{-1}$, resulting from the measured proper motion dispersion in the core of the cluster, $\sigma_\mu = 0.436 \pm 0.009$ mas yr$^{-1}$, using optical observations.

## 4. Presence of Nonluminous Matter in the Core of NGC 6752

Two pulsars in the core of the cluster, PSRB and PSRE, have a negative measured time derivative of their spin period $\dot{P}_{\rm meas}$; that is the signature of their accelerated motion in the gravitational potential well of the cluster (Paper I), since $\dot{P}$ is always intrinsically positive because of pulsar spindown. In the case of PSRD (Paper I) and PSRF (Ridolfi et al. 2021, this work) $\dot{P}_{\rm meas}$ is positive, with an absolute value at least as large as for PSRB. An insight into the Galactic population of MSPs[15] (here defined as pulsars with a spin period not larger than 10 ms) shows that it is highly unlikely that $\dot{P}_{\rm meas}$ for PSRD and PSRF is dominated by their intrinsic spindown $\dot{P}_{\rm intr}$.

In fact, the Galactic population of MSPs consists of 228 objects for which it has been measured $\dot{P} > 0$. Among them, 223 MSPs have $\dot{P} < 10^{-19}$ s s$^{-1}$, with an average spindown rate $\langle\dot{P}\rangle = 1.40 \times 10^{-20}$ s s$^{-1}$, and a standard deviation $\sigma_{\dot{P}} = 1.22 \times 10^{-20}$ s s$^{-1}$. For the four pulsars in the core, $|\dot{P}_{\rm meas}|$ is larger than $\langle\dot{P}\rangle$ by factors that range from 31 (in the case of PSRE) to 69 (in the case of PSRD). The maximum $\dot{P}$ in this sample is $7.74 \times 10^{-20}$ s s$^{-1}$ (PSR J0218+4232, Desvignes et al. 2016), a value that is an order of magnitude smaller than $\dot{P}_{\rm meas}$ of PSRF, namely the smallest value among the core pulsars for which $\dot{P}_{\rm meas} > 0$. For four of the remaining five Galactic MSPs, $\dot{P}$ is comprised between $1.05 \times 10^{-19}$ s s$^{-1}$ (PSR B1937+21, Reardon et al. 2021) and $1.63 \times 10^{-19}$ s s$^{-1}$ (PSR J1850+0242, Scholz et al. 2015). But even in the case of PSR J1850+0242, the $\dot{P}$ value is still a factor of ∼4.5 smaller than $\dot{P}_{\rm meas}$ for PSRF. Moreover, if one includes these four MSPs in the aforementioned calculation of the average spindown, one obtains $\langle\dot{P}\rangle = 1.60 \times 10^{-20}$ s s$^{-1}$ and $\sigma_{\dot{P}} = 2.00 \times 10^{-20}$ s s$^{-1}$. It is immediately seen that the $\dot{P}$ of these four MSPs is more than $4\sigma$ away from the average. The fifth one is PSR J1402+13 ($P = 5.9$ ms, $\dot{P} = 4.8 \times 10^{-17}$ s s$^{-1}$, Abdollahi et al. 2022), about which no discussion seems to be present in the literature. Because these five high $\dot{P}$ objects represent the 2.2% only of the sample of Galactic MSPs, and their $\dot{P}$ is significantly higher than the average MSPs spindown rate, we evaluated these five objects as outliers in the $\dot{P}$ distribution of the Galactic MSPs, and assumed that the distribution of the intrinsic spindown of the galactic MSPs is well represented by the 223 objects for which $0.0 < \dot{P} < 10^{-19}$ s s$^{-1}$. It could be objected that an MSP in a GC might have a $\dot{P}_{\rm intr}$ that is much higher than the average of the Galactic MSPs, since the cluster's crowded environment might have induced a different formation path for the hosted MSPs, which results in a considerably large $\dot{P}_{\rm intr}$. Nevertheless, NGC 6752 also hosts PSRA, for which Corongiu et al. (2023) obtained $\dot{P}_{\rm intr} = 5.02 \times 10^{-21}$ s s$^{-1}$, and PSRC, whose $\dot{P}_{\rm meas}$ and projected angular separation from the cluster's center of gravity imply $\dot{P}_{\rm intr} \leqslant 1.71 \times 10^{-20}$ s s$^{-1}$. Since for both PSRA and PSRC $\dot{P}_{\rm intr}$ is fully consistent with the average of the Galactic field MSPs, we can safely assume that this is also true for the pulsars in the core of NGC 6752.

The effects of the cluster environment on the core pulsars are also evident in the second time derivative of their spin periods.

In fact, the upper limits, both observational and theoretical, on the intrinsic $\ddot{P}$ for MSPs are so low, that it can safely be assumed that the intrinsic $\ddot{P}$ for the core pulsars are completely negligible in comparison to the measured ones, whose origin can thus be attributed to the cluster's dynamics.

---

[15] Data taken from the ATNF pulsar catalog psrcat (Manchester et al. 2005), catalog version 1.68 1.70, available at https://www.atnf.csiro.au/people/pulsar/psrcat/.

In Paper I it was investigated whether the acceleration imparted to the core pulsars (known at that time) could be entirely due to the observed luminous mass, or if a further amount of matter was necessary to explain their $\dot{P}_{\text{meas}}$. The authors of Paper I used the argument of the mass-to-light ratio $\mathcal{M}/\mathcal{L}_\mathcal{V}$ in the central regions (see Equation (1) in Paper I), and found that a minimum $\mathcal{M}/\mathcal{L}_\mathcal{V}$ of an order of 10, namely $\mathcal{M}/\mathcal{L}_\mathcal{V} > 9$ for PSRB and E, and $\mathcal{M}/\mathcal{L}_\mathcal{V} > 13$ for PSRD, is required to impart to these pulsars the accelerations, whose component along the line of sight produces the measured $\dot{P}$. Hereafter in this paper, in order to improve the reading of the text, the term *acceleration* will indicate the component along the line of sight of an acceleration, unless explicitly specified otherwise. These $\mathcal{M}/\mathcal{L}_\mathcal{V}$ values were much larger than the one available at that time and obtained with optical observations, $\mathcal{M}/\mathcal{L}_\mathcal{V} = 1.1$ (Pryor & Meylan 1993). The authors of Paper I calculated the amount of mass that should be present in the core of the cluster in the form of nonluminous objects, and found a lower limit of $1.3 \times 10^4 M_\odot$.

### 4.1. Our Approach

Given the high precision of our determination of both the first and second derivatives of the spin period for the core pulsars, we revisited the estimate of the nonluminous mass in the core of NGC 6752 with a different approach. Phinney (1993) detailed how $\dot{P}_{\text{meas}}$ is related to the intrinsic spindown $\dot{P}_{\text{intr}}$ and all accelerations acting on a pulsar in a GC with the following relation[16]:

$$\frac{\dot{P}_{\text{meas}}}{P} = \frac{\dot{P}_{\text{intr}}}{P} + \frac{a_{l,\text{MW}}}{c} + \frac{a_{l,\text{SHK}}}{c} + \frac{a_{l,\text{GC}}}{c} \quad (1)$$

where $c$ is the speed of light, $a_{l,\text{MW}}$ is the acceleration imparted by the Milky Way,[17] $a_{l,\text{SHK}}$ is the apparent acceleration due to the pulsar proper motion, also known as the Shklovskii (1970) effect, and $a_{l,\text{GC}}$ is the acceleration imparted by the GC. When the second derivative of the spin period is also measured, a further equation is available, that links the measured $\ddot{P}$ to the time derivative of the imparted acceleration, often referred to as the jerk. In fact, taking the time derivative of Equation (1) and considering relevant terms only, one obtains:

$$\frac{\ddot{P}_{\text{meas}}}{P} = \frac{\ddot{P}_{\text{intr}}}{P} + \frac{\dot{a}_{l,\text{GC}}}{c} + \frac{\dot{a}_{l,\text{NN}}}{c} \quad (2)$$

where we have introduced the effect of the nearest neighbor star, $\dot{a}_{l,\text{NN}}$, which cannot be in general neglected in crowded environments like GCs (see, e.g., Abbate et al. 2019), while we dropped the time derivatives of the Milky Way acceleration and that of the Shklowskii effect. These two terms can be neglected because, across the epoch range spanned by the observations, the 3D motion of the cluster in the Milky Way does not produce significant changes on its position, hence on the acceleration imparted by the Galaxy. The pulsar proper motion changes even less, hence the apparent acceleration due to its transverse motion remains substantially constant.

---
[16] In all symbols, the subscript $l$ indicates the component of the represented quantity along the line of sight.
[17] Strictly speaking, it is the difference between the components along the line of sight of the accelerations that the Milky Way imparts on the pulsar and on the solar system.

The acceleration and jerk imparted by the cluster can thus be determined by solving Equation (1) with respect to $a_{l,\text{GC}}$ and Equation (2) with respect to $\dot{a}_{l,\text{GC}}$:

$$a_{l,\text{GC}} = c\left[\frac{\dot{P}_{\text{meas}}}{P} - \frac{\dot{P}_{\text{intr}}}{P}\right] - a_{l,\text{MW}} - a_{l,\text{SHK}} \quad (3)$$

$$\dot{a}_{l,\text{GC}} = c\left[\left(\frac{\ddot{P}_{\text{meas}}}{P}\right) - \left(\frac{\ddot{P}_{\text{intr}}}{P}\right)\right] - \dot{a}_{l,\text{NN}} \quad (4)$$

and the link with the cluster structural parameters is obtained by replacing $a_{l,\text{GC}}$ in Equation (3) and $\dot{a}_{l,\text{GC}}$ in Equation (4) with the expression that results by assuming a given mass distribution for the cluster. In the simplest case of a stationary and spherically symmetric distribution for the mass of the cluster, $a_{l,\text{GC}}$ takes the well-known simple form:

$$a_{l,\text{GC}} = -\frac{GM(r)}{r^3}r_l \quad (5)$$

where $M(r)$ is the cluster mass enclosed in a sphere of radius $r$, centered at the center of gravity of the cluster, and $G$ is the Gravitational constant ($6.67 \times 10^{-11}$ m$^3$ kg$^{-1}$ s$^{-2}$). We express a generic position, with respect to the cluster center, in terms of two coordinates perpendicular to the line of sight, $r_\alpha \equiv (\alpha_{\text{PSR}} - \alpha_{\text{GC}})\cos\delta_{\text{GC}}$ and $r_\delta \equiv \delta_{\text{PSR}} - \delta_{\text{GC}}$, and $r_l$, namely the pulsar distance from the center along the line of sight; the relation $r = \sqrt{r_\alpha^2 + r_\delta^2 + r_l^2}$ obviously holds. In these definitions, $\alpha_{\text{PSR}}$ and $\delta_{\text{PSR}}$ are the celestial coordinates of a given pulsar, while $\alpha_{\text{GC}}$ and $\delta_{\text{GC}}$ are the celestial coordinates of the cluster gravity center, and $r_l$ is defined so that its value is positive for positions farther than the cluster center with respect to the observer, and negative otherwise. At this stage, we considered $r_\alpha$ and $r_\delta$ of each pulsar as known exactly, despite the uncertainties on the assumed coordinates of the center of gravity of the cluster (0.″5 in both coordinates, Ferraro et al. 2003). In Section 6 we will discuss the impact of these uncertainties on our results.

The analytic expression for $\dot{a}_{l,\text{GC}}$ is obtained by taking the time derivative of Equation (5):

$$\dot{a}_{l,\text{GC}} = \frac{d}{dt}\left(-\frac{GM(r)}{r^3}r_l\right) = -G\frac{dM(r)}{dr}\frac{\boldsymbol{r}\cdot\boldsymbol{v}}{r^4}r_l \\ - \frac{GM(r)}{r^3}\left(v_l - 3\frac{\boldsymbol{r}\cdot\boldsymbol{v}}{r^2}r_l\right) \quad (6)$$

where $\boldsymbol{v}$ is the pulsar velocity with respect to the cluster center, whose components are $v_\alpha \equiv dr_\alpha/dt = (\mu_{\alpha,\text{PSR}}\cos\delta_{\text{PSR}} - \mu_{\alpha,\text{GC}}\cos\delta_{\text{GC}})D$ and $v_\delta \equiv dr_\delta/dt = (\mu_{\delta,\text{PSR}} - \mu_{\delta,\text{GC}})D$, where in turn $D$ is the cluster distance, $\mu_{\alpha,\text{PSR}}\cos\delta_{\text{PSR}}$ and $\mu_{\delta,\text{PSR}}$ are the proper motion components of a pulsar, while $\mu_{\alpha,\text{GC}}\cos\delta_{\text{GC}}$ and $\mu_{\delta,\text{GC}}$ are the proper motion components of the cluster, and $v_l \equiv dr_l/dt$. Given the above definition of $r_l$, $v_l$ is positive when the pulsar is moving away from the observer, and negative otherwise. We did not consider the quantities $v_\alpha$ and $v_\delta$ to be exactly known, since their uncertainty is comparable to their amplitude and to the central velocity dispersion.

The problem is so far underdetermined: for a given pulsar, only two equations are available, but the unknown terms are its depth inside the cluster $r_l$, its 3D velocity $\mathbf{v}$, and all the parameters that enter in the analytic expression of a given mass distribution for the cluster. We indicate the set of all of these parameters with the formal vector $\mathbf{s} \equiv (s_1, s_2, \ldots, s_N)$. Moreover, the contribution to $\dot{a}_{l,\text{GC}}$ due to the nearest neighbor calls into play some further parameters that describe the cluster structure and dynamic properties at the pulsar position (see below).

### 4.2. The Bayesian Analysis

The availability of more than one pulsar does not allow us to fully constrain the problem, since each object has its own position and velocity, and also, the cluster's local properties differ from a given position to another. For this reason we adopted a Bayesian approach. All terms in the right-hand side of Equation (3) but the intrinsic spindown (see below) are known. We can then assume that they obey a Gaussian probability distribution with a mean value equal to their value, either measured or derived, and a standard deviation equal to their $1\sigma$ uncertainty. This in turn implies that also $a_{l,\text{GC}}$ (hereafter $a_l$ for notation simplicity) follows a Gaussian probability distribution:

$$P(a_l) = \frac{1}{\sigma_{a_{l,\text{m}}}\sqrt{2\pi}} \exp\left\{-\frac{1}{2}\frac{(a_l - a_{l,\text{m}})^2}{\sigma_{a_{l,\text{m}}}^2}\right\} \quad (7)$$

whose mean value $a_{l,\text{m}}$ is given by Equation (3), and whose standard deviation $\sigma_{a_{l,\text{m}}}$ is given by the usual rules for the propagation of the uncertainties. Once $a_l$ is substituted with the explicit expression given by Equation (5), where in turn $M(r)$ is substituted with the analytical expression that corresponds to a chosen model for the mass distribution, Equation (7) returns the probability distribution $P_{a_l,\text{X}}(\mathbf{s}, r_l)$, for the structural parameters in $\mathbf{s}$ and the still unknown pulsar depth in the cluster $r_l$, as it results by considering a single pulsar alone, identified by the formal index X (the subscript $a_l$ in the symbol $P_{a_l,\text{X}}(\mathbf{s}, r_l)$ indicates that this probability distribution comes from the analysis of the acceleration, in order to distinguish it from the distribution that results from the analysis of the jerk).

We calculated the Milky Way acceleration by applying Equation (16) in Lazaridis et al. (2009), but using for the vertical component of the Galactic acceleration $F_z$ the analytic formula provided by Li & Widrow (2021) (Equation (14)). The adopted values for the Solar motion in the Galaxy are $\Theta_\odot = 240.5 \pm 4.1$ km s$^{-1}$ and $R_\odot = 8.275 \pm 0.034$ kpc (Gravity Collaboration et al. 2021). At the cluster's distance and position ($l_g = 336.4929$ deg, $b_g = -25.628$ deg in Galactic coordinates) we obtain $a_{l,\text{MW}} = (+2.7 \pm 0.6) \times 10^{-11}$ m s$^{-2}$. The Shklovskii effect gives a contribution $a_{l,\text{SHK}} = (7.52 \pm 0.07) \times 10^{-11}(\mu/5 \text{ mas yr}^{-1})^2$ m s$^{-2}$, where $\mu$ is the proper motion amplitude, for a pulsar at the distance of NGC 6752. As already seen above, $\dot{P}_\text{intr}$ can be as high as several $10^{-20}$ s s$^{-1}$, a value that would give a $\sim 5\%$ contribution on the observed accelerations. Therefore, we calculated $a_{l,\text{m}}$ as $a_{l,\text{m}} = c\dot{P}_\text{meas}/P - a_{l,\text{SHK}} - a_{l,\text{MW}}$, and we took into account our poor knowledge of the intrinsic spindown by adding in quadrature the quantity $c/P \times (\langle \dot{P} \rangle + \sigma_{\dot{P}})$ to the uncertainty on $a_{l,\text{m}}$ that results from the standard propagation of the uncertainties on the three known terms.

We applied the method illustrated above to Equation (4), for deriving the probability distribution obeyed by $\dot{a}_{l,\text{GC}}$ (hereafter $\dot{a}_l$ for notation simplicity), which in turn translates into the probability distribution $P_{\dot{a}_l,\text{X}}$ for the structural parameters in $\mathbf{s}$ and the unknown quantities for a generic pulsar X. In this case it must be taken into account that the nearest neighbor contribution on the measured jerk cannot be calculated but only statistically treated and that, most important, it does not follow a Gaussian distribution, but a Lorentzian one (see, e.g., Abbate et al. 2019 and references therein):

$$P(\dot{a}_\text{NN}) = \frac{1}{\pi}\frac{\dot{a}_0}{\dot{a}_\text{NN}^2 + \dot{a}_0^2}. \quad (8)$$

The quantity $\dot{a}_0$ is dubbed as *characteristic jerk*, and it is related to the cluster's structure (Prager et al. 2017) by the following relation:

$$\dot{a}_0 = \frac{2\pi\xi}{3}G\sigma_\text{v} n\langle m \rangle \quad (9)$$

where $\xi \simeq 3.04$ is a numerical constant, while $\sigma_\text{v}$, $n$, and $\langle m \rangle$ are the cluster's 1D velocity dispersion, star number density and mean stellar mass at the pulsar position. The product $n\langle m \rangle$ can be set equal to the mass density at the pulsar position, while $\sigma_\text{v}$ can be only constrained to be smaller than the central 1D velocity dispersion. Once a model is assumed for the cluster's mass distribution, the probability distribution for $\dot{a}_\text{NN}$ depends on the pulsar position and the local velocity dispersion $\sigma_\text{v}$. Therefore, $P_{\dot{a}_l,\text{X}}$ depends on the parameters in $\mathbf{s}$, the pulsar depth in the cluster, the pulsar 3D velocity in the cluster frame, and the cluster's 1D velocity dispersion at the pulsar position. Because of the peculiar distribution for $\dot{a}_\text{NN}$, $P_{\dot{a}_l,\text{X}}(\mathbf{s}, r_l, \mathbf{v}, \sigma_\text{v})$ must be calculated from more general principles of statistics. Let $\alpha$ and $\beta$ be two generic quantities, which follow probability distributions $P_\alpha(\alpha)$ and $P_\beta(\beta)$ respectively, and $\gamma = \alpha - \beta$ their difference. The probability distribution for $\gamma$, $P(\gamma)$, is given by the following integral[18]:

$$P(\gamma) = \int_{-\infty}^{+\infty} P_\alpha(\alpha) P_\beta(\alpha - \gamma) d\alpha. \quad (10)$$

In our case $\alpha = c\ddot{P}/P$, $\beta$ is indeed $\dot{a}_\text{NN}$, and $\gamma$ is the value for $\dot{a}_l$ as predicted by Equation (6) once a given mass distribution is assumed. Because of the measurement of the second-order time derivative of the spin period, $\ddot{P}$ follows a Gaussian distribution with mean value $\ddot{P}_\text{meas}$, and standard deviation $\sigma_{\ddot{P}_\text{meas}}$. As a direct consequence, once $P$ is kept fixed, also $c\ddot{P}/P$ follows a Gaussian distribution with mean value $c\ddot{P}_\text{meas}/P$ and standard deviation $c\sigma_{\ddot{P}_\text{meas}}/P$. We completely neglected the contribution due to the intrinsic evolution of the spin period, namely $\ddot{P}_\text{intr}$ since, as we already commented on, it is negligible in comparison with $\ddot{P}_\text{meas}$.

The joint probability of simultaneously having an acceleration $a_l$ and a jerk $\dot{a}_l$ acting on a given pulsar X is the product of the two above-mentioned distributions. Its marginalization with respect to $r_l$, $\mathbf{v}$ and $\sigma_\text{v}$ returns the probability distribution $P_\text{X}(\mathbf{s})$ for the parameters that describe the assumed mass model:

$$P_\text{X}(\mathbf{s}) = \iiint P_{a_l,\text{X}}(\mathbf{s}, r_l) P_{\dot{a}_l,\text{X}}(\mathbf{s}, r_l, \mathbf{v}, \sigma_\text{v}) \tilde{P}(r_l)$$
$$\times \tilde{P}(\mathbf{v}) \tilde{P}(\sigma_\text{v}) dr_l\, d^3\mathbf{v}\, d\sigma_\text{v} \quad (11)$$

---

[18] One can easily demonstrate that, if $\alpha$ and $\beta$ follow Gaussian distributions centered at $\alpha_0$ and $\beta_0$ with standard deviations $\sigma_\alpha$ and $\sigma_\beta$ respectively, Equation (10) results in another Gaussian function centered on $\alpha_0 - \beta_0$, and with standard deviation $\sqrt{\sigma_\alpha^2 + \sigma_\beta^2}$, i.e., the usual result of the uncertainty propagation for the difference of two Gaussian distributed quantities.

where the $\tilde{P}$ functions are the prior probability distributions for their arguments. We assumed that $\tilde{P}(r_l)$ is flat and nonzero only for those values for $r_l$ that place a given pulsar within one core radius from the cluster center, i.e., $0 \leqslant |r_l| \leqslant \sqrt{r_c^2 - r_\perp^2}$, where we defined $r_\perp \equiv \sqrt{r_\alpha^2 + r_\delta^2}$, and whose sign is opposite to $\dot{P}_{\text{meas}}$, namely $r_l > 0$ if $\dot{P}_{\text{meas}} < 0$ and vice versa. In fact, both $a_{l,\text{MW}}$ and $a_{l,\text{SKH}}$ are positive, and their sum is so smaller than $c|\dot{P}_{\text{meas}}|/P$ for all core pulsars, that Equation (1) is satisfied only if $a_{l,\text{GC}}$ has the same sign of $\dot{P}_{\text{meas}}$, and the explicit expression for $a_{l,\text{GC}}$, as given by Equation (5), states that the sign of $r_l$ must be opposite to $a_{l,\text{GC}}$, hence to $\dot{P}_{\text{meas}}$.

We assumed $\tilde{P}(v)$ to be the product of the priors $\tilde{P}(v_\alpha)$, $\tilde{P}(v_\delta)$, and $\tilde{P}(v_l)$ for the three components of the velocity. Thanks to our measurement of the proper motion for all core pulsars, we assumed that $\tilde{P}(v_\alpha)$ and $\tilde{P}(v_\delta)$ are Gaussian distributions, whose mean and standard deviation can be obtained from their definition and the usual uncertainty propagation rules. Because no information is available about the radial velocity of the core pulsars, we assumed $\tilde{P}(v_l)$ to be flat and nonzero in the range $-50 \text{ km s}^{-1} \leqslant v_l \leqslant 50 \text{ km s}^{-1}$, where the limits are a conservative rounding of the escape velocity from the core, and we allowed both positive and negative values since it is not a priori known whether a pulsar is moving away or toward the observer in the cluster frame. Finally, we assumed $\tilde{P}(\sigma_v)$ to be flat in the range $0 \leqslant \sigma_v \leqslant \sigma_0$, since it cannot be larger than the central velocity dispersion. Its upper limit will be discussed later, since it is model dependent.

Because the same mass distribution must be responsible for the accelerations and the jerks observed in all the core pulsars, the final probability distribution $P(s)$ is given by the product of the distributions $P_X(s)$, later multiplied by the prior distribution $\tilde{P}(s)$:

$$P(s) = \left\{\prod_X P_X(s)\right\} \tilde{P}(s). \quad (12)$$

### 4.3. The Mass Models and the Distribution of the Luminous Mass

The models that we considered for the mass distribution in the core of the cluster are the sum of two components. The first one is due to the luminous mass, whose spatial distribution is described by the model that best fits to the data taken with optical observations. We assumed it was relatively well known and thus we kept it fixed. The second component is due to the nonluminous mass, and we investigated two scenarios for its distribution. In the first one we simply assumed that all the nonluminous mass is generically contained in a sphere, whose radius is smaller than the true distance of the closest pulsar to the cluster center (see Section 4.4). This scenario allowed us to obtain a first estimate of the amount of nonluminous matter in the core of the cluster. In the second scenario, instead, we assumed that an IMBH is located at the center of gravity of the cluster. This scenario differs from the first one (see Section 4.5), since the presence of an IMBH has a substantial impact on the mass distribution in the core, and it allowed us to test whether the resulting picture represents a physically meaningful description of the mass distribution in the core of the cluster.

The distribution of the luminous mass obviously plays a key role in our work. In fact, if one adopts a model that, e.g., underestimates its content, one obtains an overestimation of the nonluminous mass, and vice versa. For this reason, we paid great attention to the previous works, available in the literature, on the mass distribution of NGC 6752, and we identified a model that allows one to obtain a consistent picture of the optically observed structure of this cluster.

The most recent study of the mass surface and radial profile of NGC 6752 was reported in Baumgardt & Hilker (2018). The authors performed N-body simulations of the formation and evolution of the Galactic GCs. The results of their simulations for NGC 6752 (see their Figure E13) are consistent with the data at angular radii $\theta_\perp \gtrsim 10''$, but they clearly underestimate the observed surface density profile at smaller radii. This evidence can be considered consistent with the results obtained by Ferraro et al. (2003), who studied the surface star density profile of NGC 6752 using different models. In particular, they obtained a very good description of their data at all angular radii by using a combination of two King (1966) density profiles[19] (see text and Figure 5 in Ferraro et al. 2003), for separately describing the cluster star density in its inner and outer regions, in projection on the sky. We recall that a King profile describes the structure of a GC in terms of its mass density, whose analytic expression is given by the following equation:

$$\rho_K(r) = \rho_0 \left[1 + \left(\frac{r}{r_c}\right)^2\right]^{-\frac{3}{2}} \quad (13)$$

where $\rho_0$ is the central mass density and $r_c$ is the core radius. It is interesting to note that the transition between the two profiles occurs at $\theta_\perp \sim 10''$, i.e., at about the same angular radius at which the simulations by Baumgardt & Hilker (2018) begin to agree with the data. Since we have assumed that all core pulsars are indeed located in the core of the cluster, i.e., their true 3D distance from the center of gravity of the cluster is smaller than the core radius, which in turn is smaller than the angular radius at which the outer King profile begins to describe the data, we can safely assume that the luminous mass distribution that contributes to the acceleration imparted on the pulsars is well described by the inner King profile only. We tested whether the the double King profile by Ferraro et al. (2003) adequately describes the cluster profile at all radii, by calculating the cluster total mass that results from all the above-mentioned assumptions, and the already known cluster's parameters (see Table 3). We set the value for the core radius of the outer King profile to $r_c = 28''$, according to the results by Ferraro et al. (2003), and we normalized it by imposing that at $r = 10''$ it must return the same density given by the inner one. Finally, we integrated this two-component density profile up to the radius $r_{\text{NGC 6752}} = 12.26 \text{ pc}$, which corresponds to 4.27 projected half-light radii, thus following the prescription given by Leitinger et al. (2023; see their Section 4.1 for a discussion about this choice). We obtained a total mass of the cluster $M_{\text{TOT}} = 2.865 \times 10^5 M_\odot$, a value just a few percent higher than the value $(2.76 \pm 0.04) \times 10^5 M_\odot$ (see Table 3), thus ensuring that the double King profile adequately describes the optically observed mass profile of the cluster at all radii.

---

[19] Hereafter "King profile(s)" for simplicity.

### 4.4. A First Estimate of the Nonluminous Mass in the Core

Our first estimate of the nonluminous mass $M_{NL}$ in the cluster core is based on assuming that it is entirely contained in a sphere whose radius is smaller than the distance of the closest pulsar to the cluster center. We simplified the problem by assuming that the nonluminous mass does not perturb the distribution of the luminous matter. With these assumptions, this distribution for the nonluminous mass can be mathematically treated as a point-mass $M_{NL}$ placed at the center of gravity of the cluster. For this reason, this model will later be referred to as the point-mass model. The formal vector $s$ is thus the 1D vector $s \equiv (M_{NL})$. We assumed the prior distribution $\tilde{P}(M_{NL})$ to be flat and nonzero in the range $0 \leqslant M_{NL} \leqslant 10^4 M_\odot$. In this model we set to $10$ km s$^{-1}$ the upper limit for the local velocity dispersion that enters in the characteristic jerk expression. This value is a conservative upper limit for the measured central velocity dispersion, namely $\sigma_0 = 8.5 \pm 0.2$ km s$^{-1}$. We obtained $M_{NL} = 3.05^{+0.10}_{-0.12} \times 10^3 M_\odot$ at the $1\sigma$ level ($M_{NL} = 3.05^{+0.21}_{-0.23} \times 10^3 M_\odot$ and $M_{NL} = 3.05^{+0.35}_{-0.37} \times 10^3 M_\odot$ at the $2\sigma$ and $3\sigma$ level, respectively). This means that about 3 thousand solar masses of nonluminous mass, with a $3\sigma$ lower limit $M_{NL} \geqslant 2.68 \times 10^3 M_\odot$, are required to explain the measured accelerations and jerks acting on the core pulsars.

### 4.5. Is There an IMBH in the Core of NGC 6752?

We explored a scenario where an IMBH resides at the center of gravity of NGC 6752. The presence of an IMBH modifies the density profile of a GC, with respect to the predictions of a pure King profile, up to a radius $r_i$ dubbed as the IMBH influence radius (Baumgardt et al. 2004a and references therein). Such a density distribution is called *cusp*, since for $r \leqslant r_i$ it predicts higher values for the density than the ones predicted at the same radii by the King model that describes the density profile at $r > r_i$. In this scenario the nonluminous mass is not only due to the IMBH alone, but also to the extra mass $M_{CUSP} - M_{King}(r_i)$, namely the difference between the amounts of the cluster mass contained up to $r = r_i$, as predicted by the cusp density law ($M_{CUSP}$), and the King model that describes the luminous mass at radii $r \geqslant r_i$ ($M_{King}(r_i)$). Prager et al. (2017) explicitly obtained the analytic expression for the density profile of the cusp $\rho_C(r)$, as it results from the works by Baumgardt et al. (2004a, 2004b):

$$\rho_C(r) = \rho_i \left(\frac{r}{r_i}\right)^{-1.55}. \quad (14)$$

The influence radius $r_i$ is related to the IMBH mass $M_{BH}$ and the 1D velocity dispersion of the stars in the cluster core $\sigma_0$ by the formula (Baumgardt et al. 2004a, Equation (1)):

$$r_i = \frac{GM_{BH}}{2\sigma_0^2}. \quad (15)$$

Outside $r_i$ the mass density profile follows the standard King (1966) law (see Equation (13)). In the simulations by Baumgardt et al. (2004b) the radial density profile is a continuous function at $r_i$. Therefore the condition $\rho_C(r_i) = \rho_K(r_i) = \rho_0/[1 + (r_i/r_c)^2]^{3/2}$ must hold. Baumgardt et al. (2004a) also cited the further condition that the total mass of the cusp should not be larger than the mass of the IMBH, if the latter is heavier than a few percent of the cluster mass. From the point-mass model we obtain $M_{NL} \lesssim 3500 M_\odot$ at the $3\sigma$ level, i.e., about 1.4% of the cluster mass. Under the reasonable assumption that the true amount of nonluminous matter cannot be larger than twice the aforementioned upper limit, this constraint is not necessary in this case. A visual inspection of Figure 5 in Ferraro et al. (2003) led us to place an upper limit on $r_i$. In fact, if the cusp is present, it should leave a signature in the optically observed radial profile of the cluster. In our inspection of the aforementioned figure, we could not identify any deviation from the King profile that describes the data at angular radii up to 10″. We thus deduce that, if indeed a cusp is present in the core, its radius of influence cannot be larger than the $x$-coordinate $x_1$ of the first data point in the plot, which we extrapolated to be $\log_{10}(x_1/\text{arcseconds}) = 0.1$, i.e., 0.025 pc at the cluster distance. We conservatively assumed that the cusp can extend up to the $x$-coordinate $x_2$ of the second data point, which we extrapolated to be $\log_{10}(x_2/\text{arcseconds}) = 0.6$, i.e., 4″0, which corresponds to 0.08 pc at the cluster distance. This value is roughly equal to the projected distance of PSRB from the cluster's center of gravity, the core pulsar with the second highest projected distance. One can easily verify that in the cases of PSRB, PSRD, and PSRE, the parameter space that results from all the considerations above also contains a subspace where these pulsars are located inside the cusp, thus ensuring that these three objects can be used as probes of its structure.

The resulting cluster's mass profile $M(r)$, including the IMBH, is then:

$$M(r) = \begin{cases} M(r) = M_{BH} + 4\pi \int_0^r r^2 \rho_C(r) dr & \text{if } r < r_i \\ M(r) = M_{BH} + 4\pi \int_0^{r_i} r^2 \rho_C(r) dr + 4\pi \int_{r_i}^r r^2 \rho_K(r) dr & \text{if } r \geqslant r_i \end{cases} \quad (16)$$

where the King profile that describes the mass distribution at $r \geqslant r_i$ is the inner King profile from the studies by Ferraro et al. (2003; see Section 4.3). In this model the still unconstrained structural parameter is $M_{BH}$, hence $s = (M_{BH})$, and we assumed its prior $\tilde{P}(M_{BH})$ to be flat and nonzero in the conservative range $0 \leqslant M_{BH} \leqslant 10^4 M_\odot$. We also included $\sigma_0$ among the parameters of our probability space, with a Gaussian prior whose mean and standard deviation are its measured value and $1\sigma$ uncertainty, respectively. We obtained $M_{BH} = 2.67^{+0.05}_{-0.08} \times 10^3 M_\odot$ at the $1\sigma$ level ($2.67^{+0.12}_{-0.16} \times 10^3 M_\odot$ and $2.67^{+0.18}_{-0.24} \times 10^3 M_\odot$ at the $2\sigma$ and $3\sigma$ levels, respectively), while the overall amount of the nonluminous mass, $M_{NL} = M_{BH} + M_{CUSP} - M_{King}(r_i)$, now results in $M_{NL} = 3.08^{+0.08}_{-0.06} \times 10^3 M_\odot$ at the $1\sigma$ level ($3.08^{+0.14}_{-0.14} \times 10^3 M_\odot$ and $3.08^{+0.20}_{-0.25} \times 10^3 M_\odot$ at the $2\sigma$ and $3\sigma$ levels, respectively). Figure 2 displays the triangular plot for the posterior probability of the $M_{BH}$–$\sigma_0$ space. The black line in the color plot marks the boundary between the two regions where the observational constraint on the IMBH radius of influence is (left side) and is not (right side) satisfied. We immediately see that the points with the highest probability density lie at the boundary of the allowed region, and not inside it, and this leads us to put into question the physical validity of our result, i.e., if there is indeed a $\sim 2.7 \times 10^3 M_\odot$ IMBH at the center of the cluster, or whether it is simply a mathematical solution for our problem with no real physical meaning.

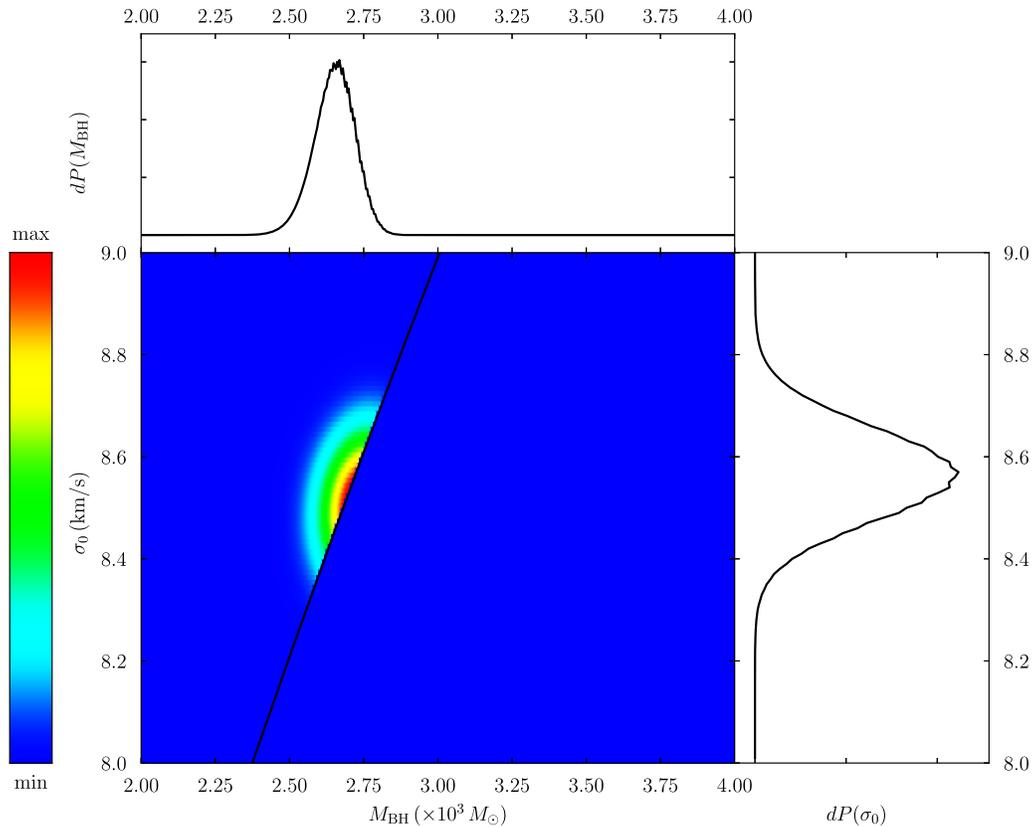

**Figure 2.** Posterior probability for the $M_{BH}$–$\sigma_0$ space resulting from the analysis aimed to constrain the possible presence of an IMBH in the core of NGC 6752. The horizontal and vertical scales have been zoomed in the ranges $2 \times 10^3 M_\odot$–$4 \times 10^3 M_\odot$ and 8 km s$^{-1}$–9 km s$^{-1}$, respectively, for a better inspection of the relevant portion of each plot. The diagonal black line in the colored plot delimits the observational constraint on the size of the IMBH influence radius. The scale for the 2D color map, displayed in the vertical left panel, is normalized to the maximum value of the 2D probability distribution. Units for all probability distributions are arbitrary.

Figure 3 also displays the posterior probability of the $M_{BH}$–$\sigma_0$ space, but in the case where the constraint on the IMBH radius of influence is not applied. The points with the highest probability are located in the not allowed region, and those points, that show in Figure 2 the highest probability, have now a probability not negligibly lower than the maximum. Quantitatively speaking, the probability that the true solution of the problem lies in the allowed region, i.e., that the IMBH scenario has a real physical meaning, is $\sim$0.25 only. This probability is not low enough for an immediate ruling out of this model, nor high enough to consider physically certain the presence of an IMBH in the core of NGC 6752. We could in principle relax the upper limit on the radius of influence, allowing $r_i$ to be larger than the $4''\!.0$ limit assumed above. This would shift, in Figure 3, the diagonal line toward larger values for $M_{BH}$, thus increasing the probability that the IMBH scenario has a real physical meaning. But a cusp with such a large radius would already have been detected in the analysis by Ferraro et al. (2003), at odds with their results.

If we, instead, put a more stringent upper limit on $r_i$, assuming that its angular size at the cluster distance is not larger than the first data point of Figure 5 in Ferraro et al. (2003), i.e., $r_i \leqslant x_1 = 1''\!.25$, further considerations arise against the physical validity of the IMBH scenario. It must be noted that, with such a tight constraint on $r_i$, the cusp radius would be smaller than the projected distance of PSRE from the cluster's gravity center. We recall here that PSRE is the core pulsar with the smallest angular distance from the cluster center. This means that none of the core pulsars would be located inside the cusp, hence none of them could probe the cusp structure. As a consequence, the IMBH model would be totally indistinguishable from the point-mass model, and it should return exactly the same results for the overall amount of the nonluminous mass. It is easy to check whether this was the case, namely if we could obtain an overall amount for the nonluminous of at least $2.66 \times 10^3 M_\odot$ (3$\sigma$ lower limit, see Section 4.4), by applying our methods with this new tighter constraint on $r_i$. If we conservatively impose that the central velocity dispersion $\sigma_0$ lies in the range 6.5–10.5 km s$^{-1}$, i.e., within 10$\sigma$ from its measured value of 8.5 km s$^{-1}$ (see Table 3), the overall a priori predicted amount of the nonluminous mass can be at most $1.3 \times 10^3 M_\odot$, in clear contradiction with the results given by the point-mass model. The IMBH model can, again a priori, predict $M_{NL} \geqslant 2.66 \times 10^3 M_\odot$, only if the cluster's central velocity dispersion is at least 15.1 km s$^{-1}$, which is too large with respect to the one inferred from the optical observations.

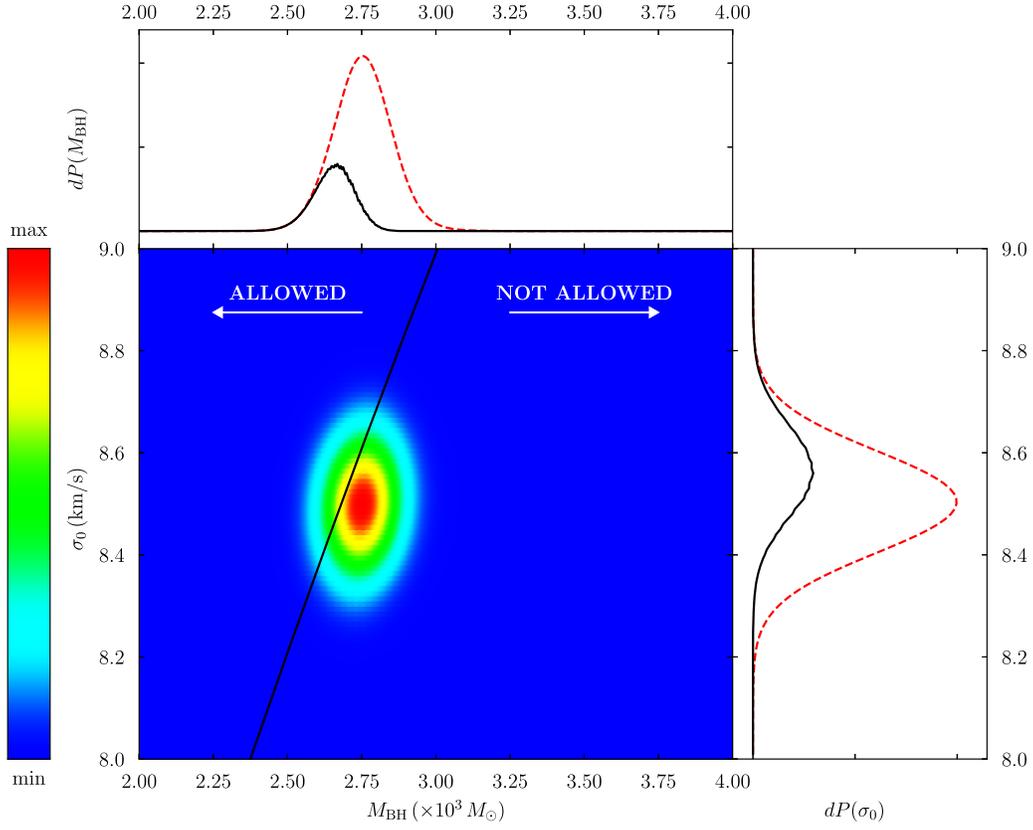

**Figure 3.** Same as Figure 2, but also considering those points in the $M_{BH}$–$\sigma_0$ space, whose corresponding value for the IMBH influence radius is larger than the upper limit we imposed on $r_i$ (see Section 4.5). Labels have been added for marking and easily identifying in which region our condition on $r_i$ is obeyed ("ALLOWED") or violated ("NOT ALLOWED"). In both the uppermost and rightmost panels, the black solid and red dashed lines are the posterior probability distributions that are obtained by considering, in the marginalization, the allowed region only and the entire $M_{BH}$–$\sigma_0$ space, respectively. All 1D posteriors have been plotted before being normalized after the marginalization, in order to highlight their relative height.

All of these considerations point against the presence in the core of the cluster of an IMBH, either with a mass equal to or higher than few thousands solar masses, or anyway massive enough to be responsible for the presence of the extra mass that explains the measured derivatives of the spin period for the core pulsars. The results from the point-mass model firmly state that an amount of nonluminous mass of at least $\sim 3 \times 10^3 M_\odot$ is necessary to justify the observed accelerations and jerks acting on the core pulsars, but it is very unlikely that the IMBH model is able to provide this amount of mass, consistently with all other optically observed features of NGC 6752. Most likely, the only possible scenario remaining invokes the presence in the cluster core of a population of much lighter nonluminous objects, whose nature and spatial distribution cannot be investigated with the present data, and which are so far undetected in optical observations and not yet predicted by models and simulations published by other authors.

## 5. The Central Mass-to-light Ratio of NGC 6752

The mass-to-light ratio ($\mathcal{M}/\mathcal{L}_V$) in GCs provides a useful link between the directly measurable emitted light and the indirectly determinable distribution of mass, thus adding elements in the investigation of the structure of this kind of stellar associations, and probing models obtained by theoretical calculations and numerical simulations.

Following the method reported in Paper I, we revisited the mass-to-light ratio argument for PSRB, D and E, also including PSRF. Equation (1) in Paper I reads:

$$\left|\frac{\dot{P}(\theta_\perp)}{P}\right| < \left|\frac{a_{l,\max}(\theta_\perp)}{c}\right| \simeq 1.1 \frac{G}{c} \frac{M_{\mathrm{cyl}}(<\theta_\perp)}{\pi D^2 \theta_\perp^2}$$

$$= 5.1 \times 10^{-18} \frac{\mathcal{M}}{\mathcal{L}_V} \left[\frac{\Sigma_V(<\theta_\perp)}{10^4 L_{V,\odot}\,\mathrm{pc}^{-2}}\right] \mathrm{s}^{-1} \quad (17)$$

where $\theta_\perp$ is the projected angular offset of the pulsar with respect to the cluster center, $a_{l,\max}(<\theta_\perp)$ is the maximum acceleration, due to the cluster potential that an object can experience at the angular offset $\theta_\perp$, $M_{\mathrm{cyl}}(<\theta_\perp)$ is the cluster mass enclosed in a cylinder of radius $\theta_\perp$ whose main axis is parallel to the line of sight and passes through the cluster center, $\Sigma_V(<\theta_\perp)$ is the cluster's average surface brightness within the angular radius $\theta_\perp$ from its center, and $L_{V,\odot}$ is the luminosity of the Sun in the V-band. Using the value $\Sigma_V = 4.025 \times 10^4 L_\odot\,\mathrm{pc}^{-2}$, assumed constant according to the work reported in Noyola & Gebhardt (2006), we obtained $\mathcal{M}/\mathcal{L}_V >$ 4.6, 5.2, 4.6, and 4.3 (here and hereafter in this section in solar units for PSRB, PSRD, PSRE, and PSRF, in the given order). The discrepancies between these values and the ones reported in Paper I for the first three pulsars, namely $\mathcal{M}/\mathcal{L}_V \gtrsim 9$ for PSRB and PSRE, and $\mathcal{M}/\mathcal{L}_V \gtrsim 13$ for PSRD, are due to our use of more recent determinations of the surface brightness profile and the coordinates for the cluster center (see Table 3). The position of the cluster center of gravity, assumed by the authors of Paper I, also led them to obtain the much larger lower limit of $1.3 \times 10^4 M_\odot$ for the mass responsible for the accelerations acting on PSRB, PSRD, and PSRE, than the one we obtained in our analysis (see Section 4).

We also derived the $\mathcal{M}/\mathcal{L}_\mathcal{V}$ values by considering the luminous mass distribution for the cluster reported and commented in Section 4. We integrated the double King density profile (see Section 4.3) along cylinders of radii equal to the angular separation of each core pulsar, and parallel to the line of sight. At first, we considered the luminous mass only, and we found $\mathcal{M}/\mathcal{L}_\mathcal{V} = 2.0$ for PSRB, PSRD, and PSRE, and 1.9 for PSRF. These values are substantially smaller than those we obtained by applying Equation (17). We then added a mass of $2680 M_\odot$ in the core, namely the $3\sigma$ lower limit on the nonluminous mass that results by applying the point-mass model, and we obtained $\mathcal{M}/\mathcal{L}_\mathcal{V} = 5.65$, 8.15, 9.31, and 4.35, thus confirming that Equation (17) may give a lower limit that is too conservative for $\mathcal{M}/\mathcal{L}_\mathcal{V}$, hence for the amount of mass responsible for the acceleration acting on a pulsar in a GC.

It is worth mentioning a plot of the radial profile of $\mathcal{M}/\mathcal{L}_\mathcal{V}$ for NGC 6752, recently published in the online catalog of the Galactic GCs (see reference "d" in Table 3). This profile was obtained by Baumgardt (2017) by fitting the observed surface brightness profile against N-body simulations of the evolution of GCs, including the formation of nonluminous objects. We extrapolated from this plot the $\mathcal{M}/\mathcal{L}_\mathcal{V}$ values at the projected distance $r_\perp$ of each core pulsar, obtaining $\mathcal{M}/\mathcal{L}_\mathcal{V} = 3.1$, 3.2, 3.3, and 2.6. These values are larger than the ones we obtained by considering the luminous mass only, yet still smaller than the ones implied by the $\dot{P}/P$ ratios and, consequently, than the ones we obtained considering our conservative lower limit on the amount of nonluminous matter. These discrepancies clearly are nonnegligible, and seem to mean that the methods used in Baumgardt (2017) still lead to an underestimation of the mass in the core of this cluster, in the form of objects that are not detectable in optical observations.

## 6. The Sources of Uncertainty in Our Results

In our Bayesian analysis we encountered three sources of uncertainty for our results. The first one is our poor knowledge of the intrinsic $\dot{P}$, for the core pulsars, which we addressed with a correction of the uncertainty in the measured accelerations. The second one resides in the treatment of the contribution due to the nearest neighbor to the time derivatives of the accelerations, a contribution that, as we already commented on, can be only statistically addressed. The third one is due to the uncertainty on the coordinates of the center of gravity of NGC 6752, which we have not taken into account so far. In this section we discuss the impact of these three sources of uncertainty on our results.

### 6.1. The Intrinsic Spindown

We treated our poor knowledge of the intrinsic spindown of the core pulsars in terms of an uncertainty to be added in quadrature to the one on all other terms on the right-hand side of Equation (3). As we already discussed in Section 4.2, we based our quantification of this additional term on the observed distribution of the measured $\dot{P}$ for the MSPs in the Galactic field. Table 4 details all contributions to the uncertainty in the acceleration due to the GC, i.e., the uncertainty on the value one obtains by calculating $a_l$ from Equation (3), in terms of the $1\sigma$ fractional uncertainty. The term due to the intrinsic spindown dominates the uncertainty on $a_l$ by 2 orders of magnitude, with respect to the Galactic acceleration and the Shklovskii effect, and by 3 orders of magnitude with respect to the uncertainty on the measured $\dot{P}$. Nevertheless, the resulting total fractional uncertainty on $a_l$ is just a few percent, with an average of 3.9%. The resulting fractional uncertainty on the nonluminous mass can be estimated as follows. Equation (5) implies that the fractional uncertainty on $a_l$ is the same as on $M_{\rm TOT}(r)$, being the latter the total amount of mass responsible for the observed acceleration, i.e., the sum of the amount of luminous mass $M_{\rm L}$ with the amount of nonluminous mass $M_{\rm NL}$. A reasonable reference value for the amount of the luminous mass can be estimated by calculating the mass predicted by the King profile we used to describe the luminous mass in the core of the cluster (see Section 4.3), within a radius $\langle r_{\rm 3D} \rangle$ equal to the average 3D distance of the core pulsars from the center of gravity. We estimated it as $\langle r_{\rm 3D} \rangle \equiv \sqrt{3/2} \langle r_{\perp,\rm PSR} \rangle = 0.663 r_c$, where $\langle r_{\perp,\rm PSR} \rangle$ is the average angular displacement of the core pulsars from the coordinates of the cluster's gravity center. We thus obtained $M_{\rm L}(\langle r_{\rm 3D} \rangle) = 460 M_\odot$. By using the value $M_{\rm NL} = 3050 M_\odot$, namely the most probable value for the amount of nonluminous mass we obtained by applying the point-mass model, it results in $M_{\rm TOT}(\langle r_{\rm 3D} \rangle) = 3510 M_\odot$, whose 3.9% error is $137 M_\odot$. Because we considered known the distribution of the luminous mass, this value represents the contribution $\sigma_{M_{\rm NL},\dot{P}}$ to the uncertainty on the nonluminous mass that is induced from the poor knowledge of the intrinsic spindown of the core pulsars.

### 6.2. The Nearest Neighbor Contribution to the Jerks

Table 5 details all contributions to the uncertainty in the jerk due to the GC potential, namely the uncertainty on the value one obtains by calculating $\dot{a}_l$ from Equation (4), again in terms of the $1\sigma$ fractional uncertainty. All columns, but the one labeled $\sigma_{\dot{a}_{\rm NN}}/\dot{a}_l$, present the same quantities as the corresponding ones in Table 4, but for the case of the jerk instead of the acceleration, and $\ddot{P}$ instead of $\dot{P}$. The contribution to the uncertainty due to the intrinsic $\ddot{P}$ is not present since, as we already showed in Section 4, this quantity is several orders of magnitude lower than the uncertainty on $\ddot{P}_{\rm meas}$, hence it can be completely neglected. The parameter $\sigma_{\dot{a}_{\rm NN}}$ quantifies the uncertainty on the contribution to the jerk due to the nearest neighbor star. We defined it so that the probability to have $|\dot{a}_{\rm NN}| \leqslant \sigma_{\dot{a}_{\rm NN}}$ is equal to the probability that a generic Gaussian distributed quantity $x$ is within $\pm 1\sigma_x$ from its mean $x_{\rm m}$, namely:

$$\frac{1}{\pi} \int_{-\sigma_{\dot{a}_{\rm NN}}}^{+\sigma_{\dot{a}_{\rm NN}}} \frac{\dot{a}_0}{\dot{a}_0^2 + \dot{a}_{\rm NN}^2} d\dot{a}_{\rm NN} = \frac{1}{\sqrt{2\pi}\,\sigma_x} \int_{x_{\rm m}-\sigma_x}^{x_{\rm m}+\sigma_x}$$
$$\times \exp\left\{-\frac{1}{2}\frac{(x-x_{\rm m})^2}{\sigma_x^2}\right\} dx \quad (18)$$

from which one obtains $\sigma_{\dot{a}_{\rm NN}} \equiv 1.837 \dot{a}_0$. We used the reference value $\dot{a}_0 = 3.2 \times 10^{-20}$ m s$^{-3}$, which results by using in Equation (9) the distance $\langle r_{\rm 3D} \rangle = 0.663 r_c$ from the cluster center, as in Section 6.1, $\sigma_{\rm V} = \sigma_{\rm V,PSR} = 8.0$ km s$^{-1}$, namely the 1D velocity dispersion of the pulsars for the local velocity dispersion at $r = \langle r_{\rm 3D} \rangle$ (see Section 3), and the approximation $n\langle m \rangle \approx \rho_{\rm K}(\langle r_{\rm 3D} \rangle)$, where $\rho_{\rm K}(r)$ is given by Equation (13).

In this case the contribution due to the nearest neighbor dominates the overall uncertainty in $\dot{a}_l$, while the uncertainty on $\ddot{P}_{\text{meas}}$ plays a nearly negligible role. The resulting contribution $\sigma_{M_{\text{NL}},\dot{a}_{\text{NN}}}$ on the uncertainty on the nonluminous mass cannot be obtained as simply as in the case of the first derivatives, but requires some additional considerations given the complexity of the expression for $\dot{a}_l$ (see Equation (6)). The most generic form of the relation between $\sigma_{\dot{a}_l,\dot{a}_{\text{NN}}}$, namely the contribution on the uncertainty on $\dot{a}_l$ due to the poor knowledge of $\dot{a}_{\text{NN}}$, and $\sigma_{M_{\text{NL}},\dot{a}_{\text{NN}}}$ can be written as[20]:

$$\sigma_{\dot{a}_l,\dot{a}_{\text{NN}}} = \left|\frac{\partial \dot{a}_l}{\partial M_{\text{NL}}}\right| \sigma_{M_{\text{NL}},\dot{a}_{\text{NN}}} \quad (19)$$

after keeping fixed to a reference value all quantities in Equation (6) but $M_{\text{NL}}$. Assuming $\sigma_{\dot{a}_l,\dot{a}_{\text{NN}}} = \eta \dot{a}_l$, where $\eta = 0.5404$ is the average of the values in the fourth column of Table 5, the contribution on the uncertainty on $M_{\text{NL}}$, due to the poor knowledge of $\dot{a}_{\text{NN}}$ only, results:

$$\sigma_{M_{\text{NL}},\dot{a}_{\text{NN}}} = \eta \left|\frac{\partial \dot{a}_l}{\partial M_{\text{NL}}}\right|^{-1} |\dot{a}_l|. \quad (20)$$

**Table 4**
Contributions to the Uncertainties on the Pulsar Accelerations due to the GC Potential

| PSR | $\|a_l\|$ (m s$^{-2}$) | $\frac{\sigma_{(c\dot{P}_{\text{meas}}/P_0)}}{a_l}$ | $\frac{c(\langle\dot{P}_{\text{intr}}\rangle + \sigma_{\dot{P}_{\text{intr}}})/P}{a_l}$ | $\frac{\sigma_{a_{\text{MW}}}}{a_l}$ | $\frac{\sigma_{a_{\text{SHK}}}}{a_l}$ | $\frac{\sigma_{a_l}}{a_l}$ |
|---|---|---|---|---|---|---|
| PSRB | $2.85 \times 10^{-8}$ | $1.93 \times 10^5$ | 0.03303 | 0.00021 | 0.00026 | 0.03303 |
| PSRD | $3.19 \times 10^{-8}$ | $0.96 \times 10^5$ | 0.02726 | 0.00019 | 0.00018 | 0.02726 |
| PSRE | $2.86 \times 10^{-8}$ | $1.02 \times 10^5$ | 0.06006 | 0.00021 | 0.00015 | 0.06006 |
| PSRF | $2.60 \times 10^{-8}$ | $3.34 \times 10^5$ | 0.03554 | 0.00023 | 0.00068 | 0.03554 |

**Table 5**
Contributions to the Uncertainties on the Pulsar Jerks due to the GC Potential

| PSR | $\|\dot{a}_l\|$ (m s$^{-3}$) | $\frac{\sigma_{(c\ddot{P}_{\text{meas}}/P_0)}}{\dot{a}_l}$ | $\frac{\sigma_{\dot{a}_{\text{NN}}}}{\dot{a}_l}$ | $\frac{\sigma_{\dot{a}_l}}{\dot{a}_l}$ |
|---|---|---|---|---|
| PSRB | $1.62 \times 10^{-19}$ | 0.02447 | 0.36258 | 0.36340 |
| PSRD | $0.47 \times 10^{-19}$ | 0.02513 | 1.24408 | 1.24434 |
| PSRE | $2.05 \times 10^{-19}$ | 0.00405 | 0.28693 | 0.28696 |
| PSRF | $2.19 \times 10^{-19}$ | 0.01179 | 0.26799 | 0.26825 |

It must be noted that the explicit expressions for $\dot{a}_l$ and $\partial \dot{a}_l/\partial M_{\text{NL}}$ also depend on the orientation, with respect to the observer, of both the pulsar position $\mathbf{r}$ and velocity $\mathbf{v}$ in the cluster's center reference frame. Because these dependencies cannot be neglected, we averaged Equation (20) over all possible orientations of the two vectors $\mathbf{r}$ and $\mathbf{v}$, thus obtaining:

$$\langle\sigma_{M_{\text{NL}},\dot{a}_{\text{NN}}}\rangle = \frac{\eta}{3}\left|r\frac{dM(r)}{dr} - 3M(r)\right|. \quad (21)$$

We used again the results from the point-mass model (see Section 4.4), and the reference position $r = \langle r_{\text{3D}}\rangle$, as above, thus obtaining $\langle\sigma_{M_{\text{NL}},\dot{a}_{\text{NN}}}\rangle = 2193 M_\odot$.

Any comparison between the values we obtained for $\sigma_{M_{\text{NL}},\dot{P}}$ and $\langle\sigma_{M_{\text{NL}},\dot{a}_l}\rangle$ must consider that the overall probability distribution for $M_{\text{NL}}$ is the product of the two distributions that result by separately considering the acceleration and jerk. One can easily see that the product of two peaked distributions, that are peaked at values that are close to each other but with different widths of the peak, results in another function that is peaked at the average value of the peaks of the two factors, with a characteristic width which is smaller than the one of both factors. Nevertheless, the value obtained for $\langle\sigma_{M_{\text{NL}},\dot{a}_{\text{NN}}}\rangle$ seems so large that one may deduce that the probability distribution obtained by considering the jerks only might not be able to tighten the uncertainties on the nonluminous mass that is obtained by considering the accelerations only. For this reason we repeated the analysis reported in Section 4.4, but considering only the accelerations. We obtained $M_{\text{NL}} = (3.105 \pm 0.137) \times 10^3 M_\odot$ at the $1\sigma$ level ($M_{\text{NL}} = 3.105^{+0.332}_{-0.313} \times 10^3 M_\odot$ and $M_{\text{NL}} = 3.105^{+1.035}_{-0.508} \times 10^3 M_\odot$ at the $2\sigma$ and $3\sigma$ level, respectively). The best value is consistent, within the uncertainties, with the one reported in Section 4.4, but in this case, the uncertainties are notably larger than the ones obtained by including the jerks. In practice, the amount of nonluminous mass is mainly determined by the accelerations, while the jerks allow a nonnegligible tightening of the uncertainties.

---

[20] Equation (19) is the application of the known rule $\sigma_q = |\partial f(q_1, q_2,...,q_N)/\partial q_1|\sigma_{q_1}$ for propagating the uncertainties between two quantities $q$ and $q_1$ related to each other by $q = f(q_1, q_2,...,q_N)$, in the case where all other quantities $q_2,...,q_N$ are kept fixed.

### 6.3. The Cluster's Gravity Center and the Final Estimate of the Amount of Nonluminous Mass

All quantitative results, reported and commented so far, rely on the assumed position for the cluster's gravity center, which we held fixed at the values indicated in Table 3 and hereafter referred to as the measured coordinates, whose $1\sigma$ uncertainties are $0\rlap{.}''5$ in both directions. Given their nonnegligible uncertainties, we investigated whether our estimation of the nonluminous mass might change if the center of gravity were not placed at the currently bona fide coordinates. We considered positions for the gravity center with an angular displacement $r_{\text{off}} \leqslant 1''$. For each position, we recomputed the amount of nonluminous matter by applying the point-mass model. In several cases, the resulting probability distribution showed more than one single peak for $M_{\text{NL}}$. For this reason we consider in this discussion the mean mass $M_{\text{NL,mean}}(\alpha, \delta)$ so defined:

$$M_{\text{NL,mean}}(\alpha, \delta) = \frac{\int M_{\text{NL}} P(M_{\text{NL}}, \alpha, \delta) dM_{\text{NL}}}{\int P(M_{\text{NL}}, \alpha, \delta) dM_{\text{NL}}} \quad (22)$$

where $P(M_{\text{NL}}, \alpha, \delta)$ is the probability density distribution obtained for the gravity center located at the celestial coordinates $(\alpha, \delta)$. We obtained values in the range $2.87 \times 10^3 M_\odot \lesssim M_{\text{NL,mean}} \lesssim 7.54 \times 10^3 M_\odot$, with mean $\langle M_{\text{NL,mean}}\rangle = 3.9 \times 10^3 M_\odot$, standard deviation $\sigma_{\langle M_{\text{NL,mean}}\rangle} = 1.0 \times 10^3 M_\odot$, and median $\tilde{M}_{\text{NL,mean}} = 3.6 \times 10^3 M_\odot$.

Figure 4 (panel (a)) displays a histogram of the obtained values, with a bin width of $200 M_\odot$. There is a clear peak around $3000 M_\odot$, in agreement with our previous result, but also a nonnegligible tail extending up to $\sim 7600 M_\odot$. Figure 4 also contains information about the $r_{\text{off}}$ values, in steps of $0\rlap{.}''25$, from which one deduces that the tail above $5000 M_\odot$ receives contributions from positions at $r_{\text{off}} \geqslant 0\rlap{.}''5$, and that the peak at $3000 M_\odot$ is mainly due to positions with an offset in the range $0\rlap{.}''25 \leqslant r_{\text{off}} \leqslant 0\rlap{.}''75$, with also relevant contributions from nearer and farther points. Such a distribution may be a consequence of our division of the exploited area in annuli, whose areas are in the ratio 1:3:5:7. In order to have a better insight into the dependence of the amount of nonluminous mass on the position of the cluster center of gravity, we plotted in Figure 5 a color map of the obtained values for $M_{\rm NL,mean}$ as a function of the position of the center of gravity. In panel (a) the obtained color map is placed in the context of the cluster core, while panel (b) shows an insight into the region under consideration. The position of the measured coordinates lies close to the northern boundary of a clear large area where $M_{\rm NL,mean}(\alpha, \delta)$ is up to $\sim 3500\, M_\odot$. Its displacement in decl. and elongation in R.A. can be seen as related to the fact that this area covers decl. around the mean of the core pulsars. The highest values are mostly required for positions toward the northwest, where one finds the farthest points from PSRD and PSRF, which are the pulsars with the highest $|\dot P|$ (PSRD) and distance (PSRF) from the measured coordinates.

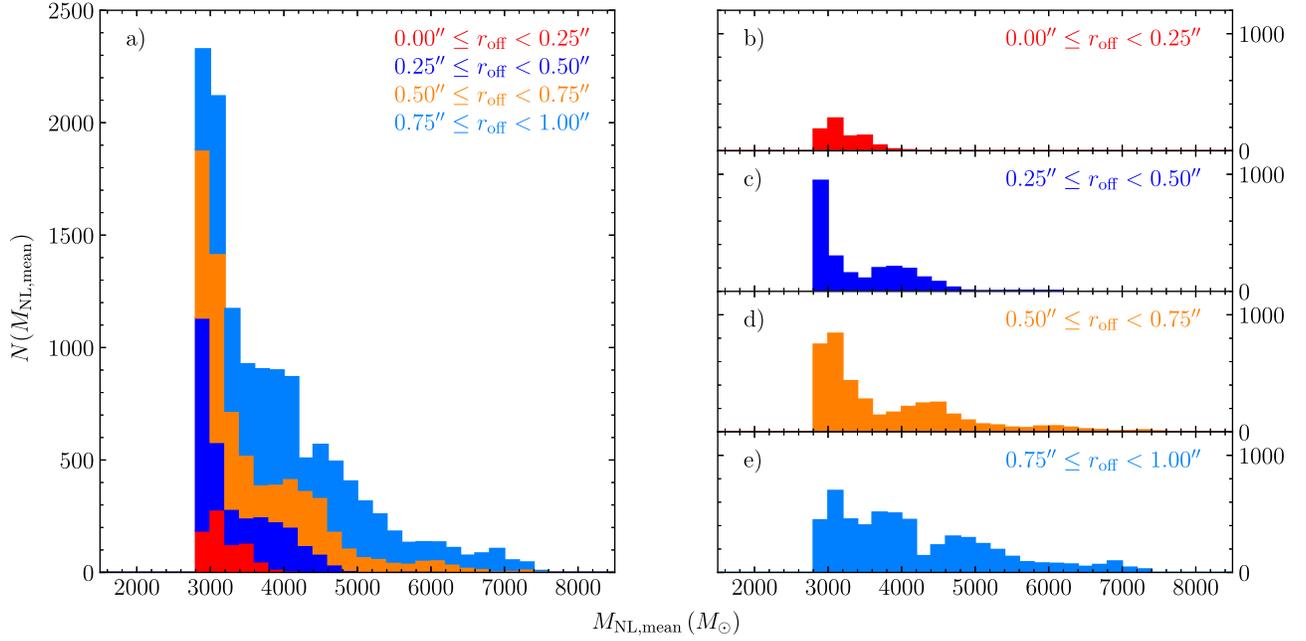

**Figure 4.** Histograms of the amount of nonluminous matter in the core of NGC 6752 for the explored positions of the cluster center of gravity. Panel (a): cumulative histogram: for each bin different colors represent the contribution at distances $r_{\rm off}$, from the coordinates of the cluster center of gravity by Ferraro et al. (2003), $r_{\rm off} < 0\farcs25$ (red), $0\farcs25 \leqslant r_{\rm off} < 0\farcs5$ (dark blue), $0\farcs5 \leqslant r_{\rm off} < 0\farcs75$ (orange), $0\farcs75 \leqslant r_{\rm off} < 1\farcs0$ (light blue), respectively. Panels (b) to (e): separate histograms for each annulus defined as above.

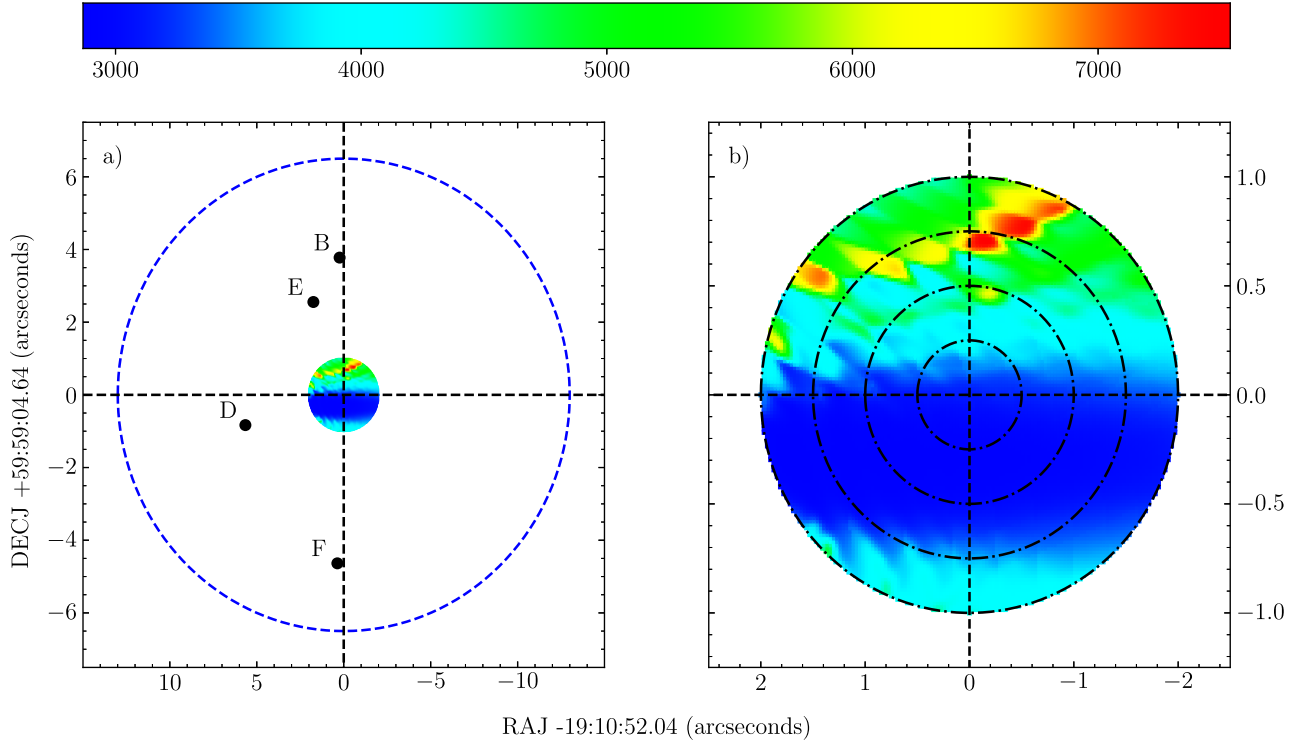

**Figure 5.** Amount of the nonluminous mass in the core of NGC 6752 as a function of the position of the cluster center of gravity. Panel (a): region containing the explored positions for the cluster center of gravity (small colored circle), compared to the position of the core pulsars (black bullets), and the core radius (blue dashed circle). Panel (b): a color map for the amount of the nonluminous mass in the core of the cluster as a function of the position of the cluster center of gravity. For each exploited position, the plotted value is the mean value of the posterior probability distribution (see Section 6.3). Dotted–dashed circles delimit the regions whose points have offsets $r_{\rm off} \leqslant 0.5\sigma$, $0.5\sigma < r_{\rm off} \leqslant 1.0\sigma$, $1.0\sigma < r_{\rm off} \leqslant 1.5\sigma$, $1.5\sigma < r_{\rm off} \leqslant 2.0\sigma$ from the innermost circle to the outermost annulus, respectively. In both panels, coordinates are centered at the measured position for the cluster center of gravity (RAJ = 19:10:52.04, DECJ = −59:59:04.64, Ferraro et al. 2003). The reference color scale is displayed in the top horizontal band.

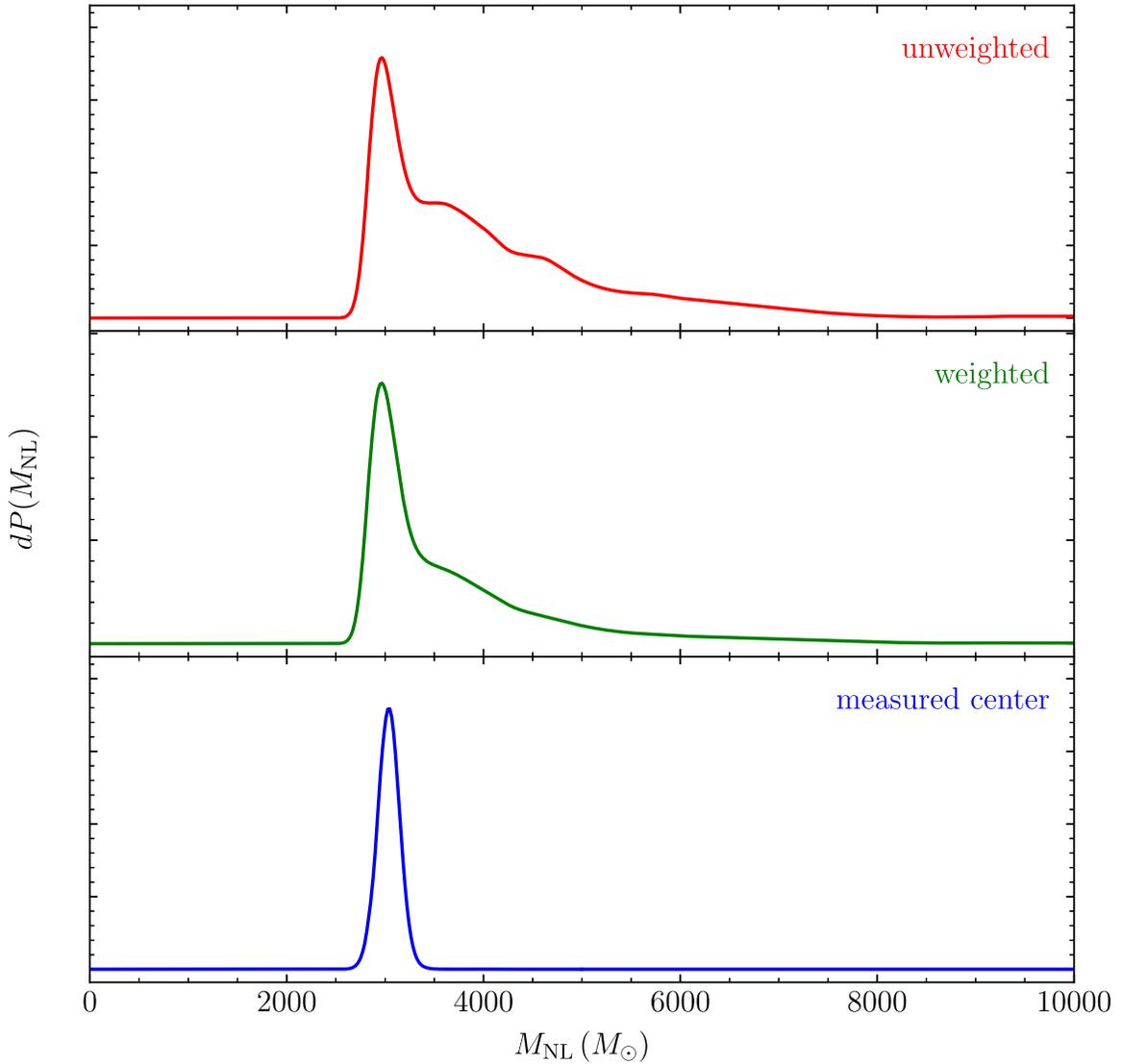

**Figure 6.** Combined unweighted (red line, top panel) and weighted (green line, middle panel) probability distributions for the nonluminous mass in the core of NGC 6752, determined as in Section 6.3. The probability distribution obtained using the coordinates of Ferraro et al. (2003) for the cluster center is also presented for immediate comparison (blue line, bottom panel, label *measured center*). Units for the probability distributions are arbitrary.

A final insight can be obtained after calculating the overall probability distribution $P(M_{\rm NL})$, by marginalizing $P(M_{\rm NL}, \alpha_{\rm GC}, \delta_{\rm GC})$ over the position of the cluster center. We performed this sum by taking into account the probability that the true center of gravity of the cluster is located in the considered position (weighted case):

$$P_{\rm W}(M_{\rm NL}) = \iint_{r_{\rm off} \leqslant 1''} P(M_{\rm NL}, \alpha_{\rm GC}, \delta_{\rm GC}) P(\alpha_{\rm GC}) P(\delta_{\rm GC}) \\ \times \cos \delta_{\rm GC}\, d\alpha_{\rm GC}\, d\delta_{\rm GC} \qquad (23)$$

where $P(\alpha_{\rm GC})$ and $P(\delta_{\rm GC})$ are Gaussian distributions with mean value $\alpha_{\rm GC}$ and $\delta_{\rm GC}$, respectively, and standard deviation $\sigma = 0\rlap{.}''5$ in both cases, and also in the case of a flat distribution for the probability density of both coordinates (unweighted case):

$$P_{\rm U}(M_{\rm NL}) = \iint_{r_{\rm off} \leqslant 1''} P(M_{\rm NL}, \alpha_{\rm GC}, \delta_{\rm GC}) \cos \delta_{\rm GC}\, d\alpha_{\rm GC}\, d\delta_{\rm GC}. \qquad (24)$$

The top and middle panels of Figure 6 display the obtained unweighted and weighted distributions, jointly with the distribution for the nonluminous mass we obtained in Section 4, i.e., in the case where the cluster center of gravity is located at the measured coordinates. Both $P_{\rm W}$ and $P_{\rm U}$ peak at the same value of $2.97 \times 10^3\, M_\odot$, and have similar lower uncertainties at $1\sigma$ ($-0.23 \times 10^3\, M_\odot$ and $-0.21 \times 10^3\, M_\odot$), at $2\sigma$ ($-0.33 \times 10^3\, M_\odot$ and $-0.29 \times 10^3\, M_\odot$, respectively) and $3\sigma$ ($-0.41 \times 10^3\, M_\odot$ and $-0.39 \times 10^3\, M_\odot$, respectively). The higher mass tails are clearly not negligible up to $8 \times 10^3\, M_\odot$ but, as shown by the histograms in Figure 4 and panel (b) in Figure 5, an amount of nonluminous mass of at least $5 \times 10^3 M_\odot$ is required only if the true gravity center of the cluster has an offset $r_{\rm off} \geqslant 0\rlap{.}''5$ from the measured coordinates, i.e., $\geqslant 1\sigma$ away from the position measured by Ferraro et al. (2003). If one extracts from the probability distribution $P_{\rm W}(M_{\rm NL})$ a measure for $M_{\rm NL}$ with its (asymmetric) uncertainties, one obtains $M_{\rm NL} = 2.97^{+0.88}_{-0.23} \times 10^3 M_\odot$ at the $1\sigma$ level, with a $3\sigma$ lower limit of $2.56 \times 10^3 M_\odot$. This is the most constrained value for $M_{\rm NL}$ we can obtain from our data after considering all the sources of uncertainty, yet its errors are much larger than the one introduced by our poor knowledge of the intrinsic first and second-order time derivative of the spin period of the core pulsars. A decrease by a factor of 10 on the uncertainties on the gravity center coordinates is still not enough for the uncertainty on $M_{\rm NL}$ to be equal to the one induced by the the poor knowledge of $\dot P$. Adopting a very conservative approach, we do not conclude with a definite value for $M_{\rm NL}$, since from both $P_{\rm W}$ and $P_{\rm U}$ we deduce a $3\sigma$ upper limit on $M_{\rm NL}$ at least of $\geqslant 10^4 M_\odot$. At the same time, we can place the solid lower limit of $M_{\rm NL} \geqslant 2.56 \times 10^3 M_\odot$ on the amount of nonluminous mass in the core of NGC 6752.

The position of the cluster's center of gravity is thus the dominant source of uncertainty on the estimation of the nonluminous mass in the core of NGC 6752. If its coordinates were known with a precision comparable to the one of the position of the pulsars in the core, the poor knowledge of both $\dot P$ and $\ddot P$ for the core pulsars would become the main sources of uncertainty. But in such a situation, the resulting uncertainty on $M_{\rm NL}$ would be of only a few percent in a case like NGC 6752, thus providing a good test bed for modeling and simulations.

## 7. Summary

We presented a timing analysis, spanning about 21 yr of observations, of the isolated pulsars PSR J1910–5959B, PSR J1910-5959C, PSR J1910–5959D, PSR J1910–5959E, and PSR J1910–5959F in the GC NGC 6752. The measured spin period derivative for the pulsars in the core is dominated by the gravitational pull of the mass in the central regions, and we estimated the amount of mass that is required to fully justify these measurements. We found that at least $2.56 \times 10^3 M_\odot$, not yet taken into account in models and/or simulations, are present under the form of nonluminous matter, within a sphere of radius not larger than the distance of the closest pulsar to the cluster center of gravity. We also explored the scenario where an IMBH resides at the center of gravity of this cluster. We found that the presence of an IMBH with a mass of at least a few thousand solar masses can explain the observed accelerations and jerks experienced by the core pulsars, but is highly, although not completely, incompatible with the observational constraints deductible by the works of other authors, and that all these constraints hardly allow this scenario to provide all the necessary extra mass. We thus inferred that this extra amount of matter is most likely in the form of a population of much lighter sub- or nonluminous objects, whose exact nature and distribution cannot be investigated with the present data. Once this extra mass is taken into account, the inferred mass-to-light ratio in the central regions is higher than that jointly deduced from optical observations and from simulations of the structure and evolution of this GC. We have also shown that the main source of uncertainty in our results is the error in the coordinates of the cluster's center of gravity, and that our poor knowledge of the intrinsic spindown of the core pulsars has a smaller impact, unless the precision on the coordinates of the cluster center were comparable to that of the core pulsars, as one can obtain from their timing.

## Acknowledgments

The MeerKAT telescope is operated by the South African Radio Astronomy Observatory, which is a facility of the National Research Foundation, an agency of the Department of Science and Innovation. SARAO acknowledges the ongoing advice and calibration of GPS systems by the National Metrology Institute of South Africa (NMISA) and the time–space reference systems department of the Paris Observatory. MeerTime data is housed on the OzSTAR supercomputer at Swinburne University of Technology supported by ADACS and the Gravitational Wave Data Centre via Astronomy Australia Ltd. MeerKAT observations used the PTUSE and APSUSE computing clusters for data acquisition, storage, and analysis. These clusters were funded and installed by the Max-Planck-Institut für Radioastronomie (MPIfR) and the Max-Planck-Gesellschaft. The Parkes radio telescope is funded by the Commonwealth of Australia for operation as a National Facility managed by CSIRO. We acknowledge the Wiradjuri people as the traditional owners of the Observatory site. M.Ba. acknowledges support from the ARC Centre of Excellence for Gravitational Wave Discovery (OzGrav) under grant CE170100004. M.Ba. acknowledges support from the Australian Research Council via grant CE170100004, the ARC Centre of Excellence for Gravitational Wave Discovery (OzGrav). Data used in this project made use of the OzSTAR supercomputer supported by ADACS and the GWDC. Part of this work has been funded using resources from the INAF Large Grant 2022 GCjewels (P.I. Andrea Possenti) approved with the Presidential Decree 30/2022. This work was also in part supported by the Italian Ministry of Foreign Affairs and International Cooperation, grant No. ZA23GR03.

### ORCID iDs

A. Ridolfi ● https://orcid.org/0000-0001-6762-2638
F. Abbate ● https://orcid.org/0000-0002-9791-7661
M. Bailes ● https://orcid.org/0000-0003-3294-3081
A. Possenti ● https://orcid.org/0000-0001-5902-3731
R. N. Manchester ● https://orcid.org/0000-0001-9445-5732
M. Kramer ● https://orcid.org/0000-0002-4175-2271
P. C. C. Freire ● https://orcid.org/0000-0003-1307-9435
M. Burgay ● https://orcid.org/0000-0002-8265-4344
F. Camilo ● https://orcid.org/0000-0002-1873-3718